\documentclass[prd,showpacs,twocolumn,
amsmath,amssymb,superscriptaddress,floatfix,nofootinbib]{revtex4-1}
\usepackage[utf8]{inputenc}
\usepackage[sort&compress]{natbib}
\usepackage[normalem]{ulem}
\usepackage{lipsum} 
\usepackage{bm}
\usepackage{times}
\usepackage{amssymb,amsbsy,amsmath,amsfonts}
\usepackage{graphicx}
\usepackage{float}
\usepackage{color}
\usepackage{morefloats}
\usepackage{rotating}
\usepackage{srcltx}
\usepackage{slashed}
\usepackage{subfigure}
\usepackage{multirow}
\usepackage{verbatim}
\usepackage{hyperref}
\usepackage{tabularx}
\usepackage{adjustbox}
\usepackage[dvipsnames]{xcolor}
\usepackage{nicefrac}
\renewcommand{\arraystretch}{1.2}

\newcommand{\zqy}[1]{{\color{blue} #1}}

\usepackage{hyperref}
\hypersetup{
	colorlinks	=true,
	urlcolor	=blue,
	linkcolor	=blue,
	citecolor	=blue,
	pdftitle	={},
	pdfauthor	={Qing-Yu Zhai, Raquel Molina, Eulogio Oset, Li-Sheng Geng},
	pdfsubject	={Study of the exotic three-body $N D^* \bar{K}^*$ system}
}

\def\GeV{\mathrm{GeV}}

\def\d{\mathrm{d}}

\begin{document}

\title{Study of the exotic three-body $N D^* \bar{K}^*$ system 
}

\author{Qing-Yu Zhai}
\affiliation{School of Physics, Beihang University, Beijing 102206, China}

\author{Raquel Molina}
\email[]{Raquel.Molina@ific.uv.es}
\affiliation{Departamento de F\'{\i}sica Te\'orica and IFIC, Centro Mixto Universidad de
Valencia-CSIC Institutos de Investigaci\'on de Paterna, Apartado 22085,
46071 Valencia, Spain}

\author{Eulogio Oset}
\email[]{Eulogio.Oset@ific.uv.es}
\affiliation{Departamento de F\'{\i}sica Te\'orica and IFIC, Centro Mixto Universidad de
Valencia-CSIC Institutos de Investigaci\'on de Paterna, Apartado 22085,
46071 Valencia, Spain}

\author{Li-Sheng Geng}
\email[]{lisheng.geng@buaa.edu.cn}
\affiliation{School of Physics, Beihang University, Beijing 102206, China}
\affiliation{Centrale Pekin, Beihang University, Beijing 100191, China}
\affiliation{Peng Huanwu Collaborative Center for Research and Education, Beihang University, Beijing 100191, China}
\affiliation{Beijing Key Laboratory of Advanced Nuclear Materials and Physics, Beihang University, Beijing 102206, China }
\affiliation{Southern Center for Nuclear-Science Theory (SCNT), Institute of Modern Physics, Chinese Academy of Sciences, Huizhou 516000, China}

\begin{abstract} 
We have studied the $N D^* \bar{K}^*$ system in the framework of the Fixed Center Approximation to the Faddeev equations, taking the exotic $D^* \bar{K}^* $ system as the cluster and allowing the N to interact with the components of the cluster. Previous studies have determined the existence of three states of spin $0,1,2$ for the $D^* \bar{K}^* $ system, the one of spin $0$ associated to the $X_0(2900)$ state observed by the LHCb collaboration. From this perspective, we find five states with total spin $1/2,3/2,5/2$, with bindings from $10$ to $30$ MeV and widths below $60$ MeV, which could be well identified. We also discuss the decay channels of these states that should help in future experimental searches of these states.   

\end{abstract}


\maketitle


\section{Introduction}

The study of three-body systems involving mesons is catching up, and much work has already been done, as reported in the review paper~\cite{MartinezTorres:2020hus}. Another relevant review paper in this direction is Ref.~\cite{Wu:2022ftm}, where an important point is stressed: the fact that, while there is no meson number conservation, which can explain why we have not observed clusters of mesons in analogy to ordinary nuclei made of baryons, where the baryon number conservation is the basic stabilizing factor, the flavor is conserved in strong interactions. Thus, mesons made of different flavors would conserve the meson number in the decays, and we could expect many clusters of meson states to be relatively stable. Work in this direction is also done in Ref.~\cite{Liu:2024uxn}. On the other hand, there is a renewed interest in new hadron states of exotic nature, which do not follow the standard pattern of $q\bar{q}$ for mesons and three quarks for baryons \cite{Esposito:2014rxa,Hosaka:2016pey,Lebed:2016hpi,Guo:2017jvc,Ali:2017jda,Olsen:2017bmm,Liu:2019zoy,Brambilla:2019esw,Chen:2022asf,Liu:2024uxn}. This proliferation of exotic states has also stimulated the search for three-body systems containing also components of exotic nature.  One such system is the $DND^*$ state, with the exotic $DD^*$ component, which has been studied in \cite{Luo:2022cun} using the Gaussian expansion method~\cite{Hiyama:2012sma,Hiyama:2003cu}, and in~\cite{Montesinos:2024eoy} using the Fixed Center Approximation, with similar results. Other similar works are the $D^*D^*D^*$ system, which has also been studied in~\cite{Luo:2021ggs,Bayar:2022bnc,Ortega:2024ecy}  and the $D^* D^* \bar{K}^*$ in~\cite{Ikeno:2022jbb},  the $DKK$ and $DK \bar{K}$ in~\cite{Debastiani:2017vhv}, the $D\bar{D}^{(*)}K$ in~\cite{Wu:2021dwy,Wu:2020job,Ren:2019bne}, or the $D^{(*)}D^{(*)}K$ in~\cite{Ren:2024mjh,Pang:2020pkl,Wu:2019vsy,MartinezTorres:2018zbl}, where references to related works can be found. 
    
In the present work, we want to study one such system, which should manifest in three different spins and with different energy in each case. The system is the $D^* \bar{K}^*N$.  In this case, the exotic component is the $D^* \bar{K}^*$. While the $DK$ interaction leads naturally to the $D_{s0}(2317)$ resonance~\cite{vanBeveren:2003kd,Barnes:2003dj,Chen:2004dy,Kolomeitsev:2003ac,Gamermann:2006nm,Guo:2006rp,Yang:2021tvc,Liu:2022dmm,Mohler:2013rwa,Lang:2014yfa,Bali:2017pdv,Cheung:2020mql,MartinezTorres:2014kpc} and $D^* K^*$ also leads to known states, as the $D_{s2}^*(2573)$~\cite{Molina:2010tx}, the  $D^* \bar{K}^*$  system has exotic nature, but it was found in Ref.~\cite{Molina:2010tx} that it produced bound states in $J=0,1,2$. A candidate for the $J=0$ state was reported by the LHCb collaboration~\cite{LHCb:2020bls,LHCb:2020pxc} by looking at the $D^-K^+$ invariant mass distribution, and named $X_0(2900)$ (now called $T_{\bar{c}\bar{s}0}$). The consistency of this assumption with the prediction of Ref.~\cite{Molina:2010tx} has been further discussed in Ref.~\cite{Molina:2020hde}, where to the light of the experimental findings of Refs.~\cite{LHCb:2020bls,LHCb:2020pxc}, more accurate predictions have been made for the states with $J=1,2$.  Support for this molecular picture has been given in Refs.~\cite{Ali:2017jda,Liu:2020nil,Huang:2020ptc,Hu:2020mxp,Chen:2020aos,Xiao:2020ltm,Kong:2021ohg,He:2020btl,Agaev:2020nrc,Abreu:2020ony,Wang:2021lwy,Ding:2024dif,Agaev:2022eyk,Yu:2023avh}, although alternative views, such as a compact tetraquark state~\cite{Karliner:2020vsi,Agaev:2020nrc,Yang:2021izl,Ozdem:2022ydv}, or a threshold effect \cite{Burns:2020epm} have also been suggested. 

We follow the molecular picture in the present work and study the $D^* \bar{K} N$ system. Taking into account that the $D^* \bar{K}^*$ subsystem is bound in $S$-wave in all the three spins $J=0,1,2$, we study this system within the Fixed Center approximation to the Faddeev equations~\cite{Chand:1962ec,Barrett:1999cw,Deloff:1999gc,Kamalov:2000iy}, assuming that the  $D^* \bar{K}^*$ forms a cluster, and the third particle, the nucleon, interacts with the cluster. The method has been widely used to study three-body systems and we refer the reader to references in Ref.~\cite{MartinezTorres:2020hus, Wu:2022ftm,Ren:2024mjh}. The cluster appears in this framework through its wave function, which can be determined from the previous study of the $D^* \bar{K}^*$ interaction in~\cite{Molina:2020hde}. Then one needs information on the $N D^*$ and $N \bar{K}^*$ interaction, which we take from the work of~\cite{Montesinos:2024eoy} and~\cite{Oset:2010tof} respectively. The application of the method to the present case gives rise to five states: one with total spin $1/2$ coming from the cluster with $J=0$, two states with total spin $1/2$ and $3/2$ coming from the cluster with $J=1$, and two states with total spin $3/2$ and $5/2$ coming from the cluster with $J=2$. We also obtain the width of the states with this procedure, and the widths obtained range from $30$~MeV to~$60$~MeV, small enough to allow the peaks obtained for $|T|^2$ to be identified. 

\section{Formalism} We assume to have a $D^*\bar{K}^*$ cluster forming each of the $J=0,1,2$ state of $D^*\bar{K}^*$ exotic molecule~\cite{Molina:2020hde}. We define two partition functions: $T_1$, collecting all diagrams where the external $N$ collides first with $D^*$, and $T_2$, collecting those where the $N$ collides first with $\bar{K}^*$. We have formally the coupled equations
\begin{align}
    \left.
    \begin{matrix}
        T_1 = t_1 + t_1\Tilde{G}_0T_2 \\
        T_2 = t_2 + t_2\Tilde{G}_0T_1 \\
    \end{matrix}
    \right\}\quad
    T = T_1 + T_2, 
    \label{coupled equations}
\end{align}
where $t_1$ and $t_2$ are the amplitudes for collision of $N$ with $D^*$ and $\bar{K}^*$ respectively, and $\tilde{G}_0$ corresponds to the $N$ propagator modulated by the wave function of the $D^*\bar{K}^*$ system.
\begin{figure}[htpb]
    \centering
    \includegraphics[width=8.5cm]{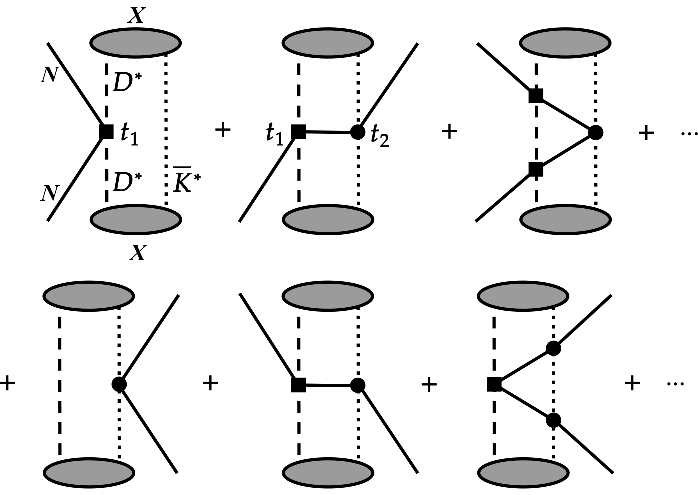}
    \caption{}
    \label{}
\end{figure}
Considering the different normalization in our meson and baryon fields, and following the steps in Refs.~\cite{Roca:2010tf,Xiao:2011rc}, we can write immediately
\begin{align}
    T=\frac{\Tilde{t}_1+\Tilde{t}_2+2\Tilde{t}_1\Tilde{t}_2\Tilde{G}_0}{1-\Tilde{t}_1\Tilde{t}_2\Tilde{G}^2_0},
    \label{T amplitude}
\end{align}
with
\begin{align}
    \Tilde{t}_1=\frac{m_X}{m_{D^*}}t_1(D^*N);\ \ \Tilde{t}_2=\frac{m_X}{m_{K^*}}t_2(\bar{K}^*N),
\end{align}
with $X$ standing for the $D^*\bar{K}^*$ cluster, and
\begin{align}
    \Tilde{G}_0=\frac{1}{2m_X}\int\frac{\d^3q}{(2\pi)^3}\ \frac{m_N}{E_N(\pmb{q})}\frac{1}{q^0-E_N(\pmb{q})+i\epsilon}F_X(q),
    \label{G0}
\end{align}
where $E_N(\pmb{q})=\sqrt{m_N^2+\pmb{q}^2}$, $q^0$ is the energy of the $N$ in the rest frame of $D^*\bar{K}^*$, given by~\cite{Bayar:2022bnc}
\begin{align}
    q^0=\frac{s-m_N^2-m_X^2}{2m_X},
    \label{q0}
\end{align}
and $F_X(q)$ is the form factor of the $D^*\bar{K}^*$ state, which is defined as~\cite{Roca:2010tf}
\begin{align}
    F_X(q)=\int\d^3r\ \psi^*(r)\psi(r)e^{-i\pmb{q}\cdot\pmb{r}}. 
    \label{definition of th eform factor}
\end{align}
Writing this form factor in terms of the wave function in momentum space is convenient. After the calculation, the form factor is given by~\cite{MartinezTorres:2020hus}
\begin{widetext}
\begin{align}
    F_X(q)=\frac{1}{\mathcal{N}}\int_{|\pmb{p}|\leq q_\mathrm{max},|\pmb{p}-\pmb{q}|\leq q_\mathrm{max}}\frac{\d^3p}{(2\pi)^3}\ \frac{1}{m_X-E_{D^*}(\pmb{p})-E_{\bar{K}^*}(\pmb{p})}\frac{1}{m_X-E_{D^*}(\pmb{p}-\pmb{q})-E_{\bar{K}^*}(\pmb{p}-\pmb{q})},
    \label{form factor}
\end{align}
\end{widetext}
where $E_{D^*}(\pmb{p})=\sqrt{m_{D^*}^2+\pmb{p}^2}$, $E_{\bar{K}^*}(\pmb{p})=\sqrt{m_{\bar{K}^*}^2+\pmb{p}^2}$ and $\mathcal{N}$ is the normalization factor, which sets the form factor normalized to unity when $\vec{q}=0$
\begin{align}
    \mathcal{N}=\int_{|\pmb{p}|\leq q_\mathrm{max}}\frac{\d^3p}{(2\pi)^3}\ \left(\frac{1}{m_X-E_{D^*}(\pmb{p})-E_{\bar{K}^*}(\pmb{p})}\right)^2.
    \label{normalization factor}
\end{align}
In Eq.~(\ref{form factor}), we have introduced a cutoff $q_\mathrm{max}$, which is the regulator used in the loops to get the pole in the $D^*\bar{K}^*$ interaction. We take $q_\mathrm{max}=1.1\,\GeV$ suited to get the \zqy{$T_{cs}$} state in the work of Ref.~\cite{Molina:2020hde}. 

We now briefly discuss the spin and isospin structure of the amplitudes. Following Refs.~\cite{Bayar:2022bnc,Roca:2010tf,Xiao:2011rc}, we have for isospin,
\begin{equation} \label{isospin}
    \begin{split}
        t_1=\frac{3}{4}t_{ND^*}^{I=1}+\frac{1}{4}t_{ND^*}^{I=0},\\
        t_2=\frac{3}{4}t_{N\bar{K}^*}^{I=1}+\frac{1}{4}t_{N\bar{K}^*}^{I=0}.
    \end{split}
\end{equation}
We only consider the $I=0$ channel in the $ND^*$ and $N\bar{K}^*$ subsystems and neglect the contribution of the $I=1$ part. Note that although there probably exists $\Sigma^*(1430)$ as a $I=1$ state of $N\bar{K}$~\cite{Belle:2022ywa}, the strength of its amplitude is very small compared to that of  the $I=0$ channel~\cite{Roca:2013cca} (see also the very small strength in the Belle experimental paper~\cite{Belle:2022ywa}). Similarly, the $N\bar{K}^*$, $I=1$ state generated in Ref.~\cite{Oset:2010tof} has a very small strength compared to that of the $I=0$ state. Thus we take $t_1=\frac{1}{4}t_{ND^*}^{I=0}$, $t_2=\frac{1}{4}t_{N\bar{K}^*}^{I=0}$. Concerning the spin structure, since the spin of the cluster can be $J_\mathrm{clu}=0,1,2$ and the external $N$ carries spin $1/2$, the total spin of the $D^*\bar{K^*}N$ system can be $J_\mathrm{tot}=1/2,3/2$ or $5/2$. We follow here the steps of \cite{Ikeno:2022jbb} and show the spin structure of $t_1$ with different $J_\mathrm{clu}$ and $J_\mathrm{tot}$ in Table~\ref{tab:spin structures}. The spin structure of $t_2$ is the same as $t_1$, with $ND^*\rightarrow N\bar{K}^*$. We also follow the prescription that the bound state of $DN^*$ with spin $1/2$ corresponds to the $\Lambda_c(2910)$ resonance and $DN^*$ with spin $3/2$ is the $\Lambda_c(2940)$, as assumed in~\cite{Montesinos:2024eoy} and~\cite{Yue:2024paz}. The association of the $\Lambda_c(2940)$ to a $D^*N$ state has been also suggested in several works~\cite{Luo:2022cun,He:2006is,Dong:2010xv,He:2010zq,Ortega:2013fta,Wang:2020dhf,Kong:2024scz}.
\begin{table}[htpb]
 \renewcommand{\arraystretch}{2}
 \setlength{\tabcolsep}{0.3cm}
    \centering
    \caption{Spin structures of different $J_\mathrm{clu}$ and $J_\mathrm{tot}$}\label{tab:spin structures}
    \begin{tabular}{ c c c }
        \hline\hline
        $J_\mathrm{tot}$ & $J_\mathrm{clu}$ & spin structure\\
        \hline
        $1/2$ & 0 & $t_1=\frac{2}{3}t_{ND^*}^{S=3/2}+\frac{1}{3}t_{ND^*}^{S=1/2}$ \\
        $1/2$ & 1 & $t_1=\frac{1}{3}t_{ND^*}^{S=3/2}+\frac{2}{3}t_{ND^*}^{S=1/2}$ \\
        $3/2$ & 1 & $t_1=\frac{5}{6}t_{ND^*}^{S=3/2}+\frac{1}{6}t_{ND^*}^{S=1/2}$ \\
        $3/2$ & 2 & $t_1=\frac{1}{6}t_{ND^*}^{S=3/2}+\frac{5}{6}t_{ND^*}^{S=1/2}$ \\
        $5/2$ & 2 & $t_1=t_{ND^*}^{S=3/2}$ \\
        \hline\hline
    \end{tabular}
\end{table}

The $ND^*$ amplitude is approximated by
\begin{equation} \label{ND*amplitude}
    \begin{split}
        t_1^{I=0,S=1/2}(ND^*)=\frac{\left(g_{ND^*}^{I=0,S=1/2}\right)^2}{\sqrt{s_1}-m_{\Lambda_c(2910)}+i\frac{\Gamma_{\Lambda_c(2910)}}{2}}, \\
        t_1^{I=0,S=3/2}(ND^*)=\frac{\left(g_{ND^*}^{I=0,S=3/2}\right)^2}{\sqrt{s_1}-m_{\Lambda_c(2940)}+i\frac{\Gamma_{\Lambda_c(2940)}}{2}},
    \end{split}
\end{equation}
for which we take
\begin{align}
    &m_{\Lambda_c(2910)}=2910~\mathrm{MeV},\ \ \Gamma_{\Lambda_c(2910)}=51.8~\mathrm{MeV},\notag\\
   & m_{\Lambda_c(2940)}=2940~\mathrm{MeV},\ \ \Gamma_{\Lambda_c(2940)}=20~\mathrm{MeV}.\notag
\end{align}
We also need the $N\bar{K}^*$ amplitude, which is similar to the $ND^*$ amplitude
\begin{align}
    t_2^{I=0,S=1/2,3/2}(N\bar{K}^*)=\frac{\left(g_{N\bar{K}^*}^{I=0,S=1/2,3/2}\right)^2}{\sqrt{s_2}-m_{\Lambda(1800)}+i\frac{\Gamma_{\Lambda(1800)}}{2}},
    \label{NKbar*amplitude}
\end{align}
where we use the same mass $m_{\Lambda(1800)}=1800~\mathrm{MeV}$ and width $\Gamma_{\Lambda(1800)}=205.0~\mathrm{MeV}$ of $\Lambda(1800)$ for amplitudes of different spin according to Ref.~\cite{Oset:2010tof}. 

As for $g_{ND^*}$, we get it from the $S$-wave Weinberg compositeness condition~\cite{PhysRev.137.B672} using the formula of Ref.~\cite{Gamermann:2009uq} suited to the normalization of Eq.~(\ref{ND*amplitude}) and Eq.~(\ref{NKbar*amplitude}) as
\begin{align}
    g_{ND^*}^2=\frac{m_{\Lambda_c(ND^*)}}{4m_N\mu}16\pi\gamma,\ \ \gamma=\sqrt{2\mu B},
\end{align}
with $\mu$ is the reduced mass of $ND^*$, and $B$ is the corresponding ``binding energy" with respect to its threshold. We finally obtain
\begin{align}
    g_{ND^*}^{I=0,S=1/2}=3.71,\ \ g_{ND^*}^{I=0,S=3/2}=2.63.\notag
\end{align}
For $g_{N\bar{K}^*}$, we directly use $g_{N\bar{K}^*}^{I=0}=3.3$ given in Ref.~\cite{Oset:2010tof}, which can give the pole position of $\Lambda(1800)$. 

Finally we need $\sqrt{s_1}$ in Eq.~(\ref{ND*amplitude}) and $\sqrt{s_2}$ in Eq.~(\ref{NKbar*amplitude}). They are the invariant masses of the $ND^*$ and $N\bar{K}^*$ subsystems, given by~\cite{Montesinos:2024eoy}
\begin{equation} \label{s1 and s2}
    \begin{split}
        s_1=m_N^2+(\xi m_{D^*})^2+2\xi m_{D^*}q^0, \\
        s_2=m_N^2+(\xi m_{\bar{K}^*})^2+2\xi m_{\bar{K}^*}q^0,
    \end{split}
\end{equation}
where $\xi=m_X/(m_{D^*}+m_{\bar{K}^*})$ is a correction factor, which is determined by assuming the ``binding energy" of the resonance to be divided between $D^*$ and $\bar{K}^*$ proportional to their masses, and $q^0$ is the energy of the $N$ in the rest frame of $D^*\bar{K}^*$, as given in Eq.~(\ref{q0}). 

\section{Results and discussions.}

\begin{figure}[htpb]
    \centering
    \includegraphics[width=8.5cm]{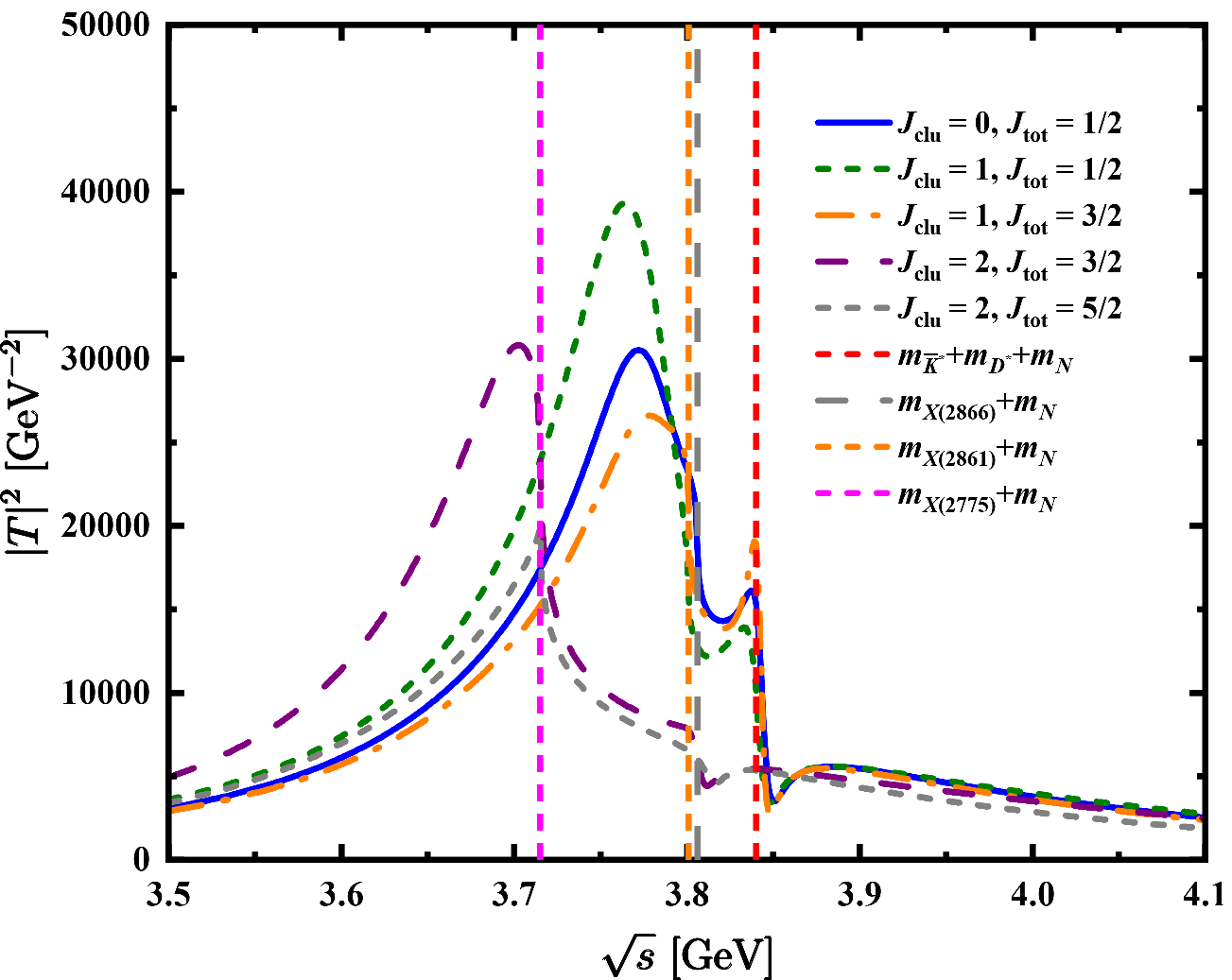}
    \caption{$|T|^2$ as a function of the total energy of the three-body system $\sqrt{s}$, for $q_\mathrm{max}=1.1\mathrm{GeV}$, assuming the $D^*\bar{K}^*$ cluster pair to be bound into the $X(2866)$, $X(2861)$, and $X(2775)$ for spin 0, 1, and 2, respectively. }
    \label{fig:tmat}
\end{figure}

We show in Fig.~\ref{fig:tmat} our results for $|T|^2$. When the cluster has $J_{clu}=0$, the total spin is $J=1/2$. In this case, we find a clear peak bound by about $25$ MeV to the $m_X + m_N$ threshold. The apparent width is about $60$ MeV. In the case of $J_{clu}=1$ and $J=1/2,3/2$, we find two peaks of similar strength, somewhat bigger than in the former case, with a binding of about $30$ MeV to the same threshold and a similar width. For $J_{clu}=2$ and $J=3/2,5/2$, we also find peaks of similar strength to the $J_\mathrm{clu}=0$ case but with smaller binding and width, particularly in the case of $J=5/2$, which has a binding of about $10$ MeV and a width of about $30$ MeV. The determination of the width of the states is a welcome feature of the FCA. We should note that elaborate frameworks, like the use of the Gaussian expansion method ~\cite{Luo:2022cun}, produce binding energies but not widths. The width of the states is important to determine the feasibility of an experiment to observe such states.
   
It is interesting to note that we also find some peaks between the two thresholds, $m_X+m_N$ and $M_{\bar{K}^*}+m_{D^*}+ m_N$. Similar peaks have also been obtained in~\cite{Montesinos:2024eoy} in the study of the $D N D^*$ system, and particularly in the study of the  $D^*D^*D^*$ system in~\cite{Ortega:2024ecy}, which are associated to states related to the Efimov effect.
   
The structure of all these states is clearly visible and should show up in devoted experiments. Certainly, the states should not be searched for in the $N \bar{K}^*\bar{D}^*$ component of these three-body systems because the system is bound for these particles. One should look into the decay channels of the subsystems. In this sense, we must note that the $D^* K^*$ cluster with $J_\mathrm{clu}=0$ can decay to $D K$. The one with $J_\mathrm{clu}=1$ can decay to $D^* K$, the one with $J_\mathrm{clu}= 2$ can decay to $DK$ and $D^* K$. Similarly, the $N D^*$ state can decay to $ND$ and the $N K^*$ state to $N K$.  Altogether, we can see that the decay channels of the three-body system would be $DKN$, $D^* KN$, and $DK^* N$. These are the channels where the peaks would show up. Peaks in invariant masses of three particles start to be seen in decays of heavy mesons into four particles in BESIII \cite{talkbeihang}. With this perspective, we are looking forward to future experimental developments where the states that we have predicted could be observed.

\section{Summary and outlook}
We have studied possible bound states of the $N D^*\bar{K}^*$ system and have found five states with different total spin, $J=1/2,3/2,5/2$ attached to bound states of the exotic  $\bar{K}^*D^*$ system, where previously three states with $J_\mathrm{clu}=0,1,2$ had been found, the one with $J_\mathrm{clu}=0$ corresponding to the $X_0(2900) $ discovered by the LHCb collaboration. We have used the Fixed Center Approximation to the Faddeev equations that has proved accurate in determining the masses and widths of some states already found, like the $\bar{K} NN$ state~\cite{Sekihara:2016vyd} seen in the in-flight $^3He(K^{-}$, $\Lambda p)n$ reaction at J-PARC~\cite{J-PARCE15:2016esq}. The states obtained have binding energies, widths, and landshapes, making them perfectly observable. The fact that some two-body subsystems correspond to exotic mesonic states makes these three-body states particularly attractive. Given the increasing experimental interest in the search for peaks in the invariant mass of three particles, we encourage the actual realization of suited experiments to see these predicted states, for which we have also discussed the decay channels where they could be observed.  

\section*{Acknowledgments.} 
L. S. G. acknowledges support from the National Key R\&D Program of China under Grant No. 2023YFA1606700 and the Natural Science Foundation of Chian under Grant No. 12435007.
R. M. acknowledges support from the CIDEGENT program with Ref. CIDEGENT/2019/015 and the PROMETEU program with Ref. CIPROM/2023/59, of the Generalitat Valenciana, and also from the Spanish Ministerio de Economia y Competitividad and European Union (NextGenerationEU/PRTR) by the grant with Ref. CNS2022-13614.  This work is also partly supported by the Spanish Ministerio de Economia y Competitividad (MINECO) and European FEDER funds under Contracts No. FIS2017-84038-C2-1-P B, PID2020-112777GB-I00, and by Generalitat Valenciana under contract PROMETEO/2020/023. This project has received funding from the European Union Horizon 2020 research and innovation program under the program H2020-INFRAIA-2018-1, grant agreement No. 824093 of the STRONG-2020 project.
\bibliography{cite}

\begin{thebibliography}{83}%
\makeatletter
\providecommand \@ifxundefined [1]{%
 \@ifx{#1\undefined}
}%
\providecommand \@ifnum [1]{%
 \ifnum #1\expandafter \@firstoftwo
 \else \expandafter \@secondoftwo
 \fi
}%
\providecommand \@ifx [1]{%
 \ifx #1\expandafter \@firstoftwo
 \else \expandafter \@secondoftwo
 \fi
}%
\providecommand \natexlab [1]{#1}%
\providecommand \enquote  [1]{``#1''}%
\providecommand \bibnamefont  [1]{#1}%
\providecommand \bibfnamefont [1]{#1}%
\providecommand \citenamefont [1]{#1}%
\providecommand \href@noop [0]{\@secondoftwo}%
\providecommand \href [0]{\begingroup \@sanitize@url \@href}%
\providecommand \@href[1]{\@@startlink{#1}\@@href}%
\providecommand \@@href[1]{\endgroup#1\@@endlink}%
\providecommand \@sanitize@url [0]{\catcode `\\12\catcode `\$12\catcode `\&12\catcode `\#12\catcode `\^12\catcode `\_12\catcode `\%12\relax}%
\providecommand \@@startlink[1]{}%
\providecommand \@@endlink[0]{}%
\providecommand \url  [0]{\begingroup\@sanitize@url \@url }%
\providecommand \@url [1]{\endgroup\@href {#1}{\urlprefix }}%
\providecommand \urlprefix  [0]{URL }%
\providecommand \Eprint [0]{\href }%
\providecommand \doibase [0]{http://dx.doi.org/}%
\providecommand \selectlanguage [0]{\@gobble}%
\providecommand \bibinfo  [0]{\@secondoftwo}%
\providecommand \bibfield  [0]{\@secondoftwo}%
\providecommand \translation [1]{[#1]}%
\providecommand \BibitemOpen [0]{}%
\providecommand \bibitemStop [0]{}%
\providecommand \bibitemNoStop [0]{.\EOS\space}%
\providecommand \EOS [0]{\spacefactor3000\relax}%
\providecommand \BibitemShut  [1]{\csname bibitem#1\endcsname}%
\let\auto@bib@innerbib\@empty
\bibitem [{\citenamefont {Martinez~Torres}\ \emph {et~al.}(2020)\citenamefont {Martinez~Torres}, \citenamefont {Khemchandani}, \citenamefont {Roca},\ and\ \citenamefont {Oset}}]{MartinezTorres:2020hus}%
  \BibitemOpen
  \bibfield  {author} {\bibinfo {author} {\bibfnamefont {A.}~\bibnamefont {Martinez~Torres}}, \bibinfo {author} {\bibfnamefont {K.~P.}\ \bibnamefont {Khemchandani}}, \bibinfo {author} {\bibfnamefont {L.}~\bibnamefont {Roca}}, \ and\ \bibinfo {author} {\bibfnamefont {E.}~\bibnamefont {Oset}},\ }\href {\doibase 10.1007/s00601-020-01568-y} {\bibfield  {journal} {\bibinfo  {journal} {Few Body Syst.}\ }\textbf {\bibinfo {volume} {61}},\ \bibinfo {pages} {35} (\bibinfo {year} {2020})},\ \Eprint {http://arxiv.org/abs/2005.14357} {arXiv:2005.14357 [nucl-th]} \BibitemShut {NoStop}%
\bibitem [{\citenamefont {Wu}\ \emph {et~al.}(2022)\citenamefont {Wu}, \citenamefont {Pan}, \citenamefont {Liu},\ and\ \citenamefont {Geng}}]{Wu:2022ftm}%
  \BibitemOpen
  \bibfield  {author} {\bibinfo {author} {\bibfnamefont {T.-W.}\ \bibnamefont {Wu}}, \bibinfo {author} {\bibfnamefont {Y.-W.}\ \bibnamefont {Pan}}, \bibinfo {author} {\bibfnamefont {M.-Z.}\ \bibnamefont {Liu}}, \ and\ \bibinfo {author} {\bibfnamefont {L.-S.}\ \bibnamefont {Geng}},\ }\href {\doibase 10.1016/j.scib.2022.08.007} {\bibfield  {journal} {\bibinfo  {journal} {Sci. Bull.}\ }\textbf {\bibinfo {volume} {67}},\ \bibinfo {pages} {1735} (\bibinfo {year} {2022})},\ \Eprint {http://arxiv.org/abs/2208.00882} {arXiv:2208.00882 [hep-ph]} \BibitemShut {NoStop}%
\bibitem [{\citenamefont {Liu}\ \emph {et~al.}(2024)\citenamefont {Liu}, \citenamefont {Pan}, \citenamefont {Liu}, \citenamefont {Wu}, \citenamefont {Lu},\ and\ \citenamefont {Geng}}]{Liu:2024uxn}%
  \BibitemOpen
  \bibfield  {author} {\bibinfo {author} {\bibfnamefont {M.-Z.}\ \bibnamefont {Liu}}, \bibinfo {author} {\bibfnamefont {Y.-W.}\ \bibnamefont {Pan}}, \bibinfo {author} {\bibfnamefont {Z.-W.}\ \bibnamefont {Liu}}, \bibinfo {author} {\bibfnamefont {T.-W.}\ \bibnamefont {Wu}}, \bibinfo {author} {\bibfnamefont {J.-X.}\ \bibnamefont {Lu}}, \ and\ \bibinfo {author} {\bibfnamefont {L.-S.}\ \bibnamefont {Geng}},\ }\href@noop {} {\  (\bibinfo {year} {2024})},\ \Eprint {http://arxiv.org/abs/2404.06399} {arXiv:2404.06399 [hep-ph]} \BibitemShut {NoStop}%
\bibitem [{\citenamefont {Esposito}\ \emph {et~al.}(2015)\citenamefont {Esposito}, \citenamefont {Guerrieri}, \citenamefont {Piccinini}, \citenamefont {Pilloni},\ and\ \citenamefont {Polosa}}]{Esposito:2014rxa}%
  \BibitemOpen
  \bibfield  {author} {\bibinfo {author} {\bibfnamefont {A.}~\bibnamefont {Esposito}}, \bibinfo {author} {\bibfnamefont {A.~L.}\ \bibnamefont {Guerrieri}}, \bibinfo {author} {\bibfnamefont {F.}~\bibnamefont {Piccinini}}, \bibinfo {author} {\bibfnamefont {A.}~\bibnamefont {Pilloni}}, \ and\ \bibinfo {author} {\bibfnamefont {A.~D.}\ \bibnamefont {Polosa}},\ }\href {\doibase 10.1142/S0217751X15300021} {\bibfield  {journal} {\bibinfo  {journal} {Int. J. Mod. Phys. A}\ }\textbf {\bibinfo {volume} {30}},\ \bibinfo {pages} {1530002} (\bibinfo {year} {2015})},\ \Eprint {http://arxiv.org/abs/1411.5997} {arXiv:1411.5997 [hep-ph]} \BibitemShut {NoStop}%
\bibitem [{\citenamefont {Hosaka}\ \emph {et~al.}(2016)\citenamefont {Hosaka}, \citenamefont {Iijima}, \citenamefont {Miyabayashi}, \citenamefont {Sakai},\ and\ \citenamefont {Yasui}}]{Hosaka:2016pey}%
  \BibitemOpen
  \bibfield  {author} {\bibinfo {author} {\bibfnamefont {A.}~\bibnamefont {Hosaka}}, \bibinfo {author} {\bibfnamefont {T.}~\bibnamefont {Iijima}}, \bibinfo {author} {\bibfnamefont {K.}~\bibnamefont {Miyabayashi}}, \bibinfo {author} {\bibfnamefont {Y.}~\bibnamefont {Sakai}}, \ and\ \bibinfo {author} {\bibfnamefont {S.}~\bibnamefont {Yasui}},\ }\href {\doibase 10.1093/ptep/ptw045} {\bibfield  {journal} {\bibinfo  {journal} {PTEP}\ }\textbf {\bibinfo {volume} {2016}},\ \bibinfo {pages} {062C01} (\bibinfo {year} {2016})},\ \Eprint {http://arxiv.org/abs/1603.09229} {arXiv:1603.09229 [hep-ph]} \BibitemShut {NoStop}%
\bibitem [{\citenamefont {Lebed}\ \emph {et~al.}(2017)\citenamefont {Lebed}, \citenamefont {Mitchell},\ and\ \citenamefont {Swanson}}]{Lebed:2016hpi}%
  \BibitemOpen
  \bibfield  {author} {\bibinfo {author} {\bibfnamefont {R.~F.}\ \bibnamefont {Lebed}}, \bibinfo {author} {\bibfnamefont {R.~E.}\ \bibnamefont {Mitchell}}, \ and\ \bibinfo {author} {\bibfnamefont {E.~S.}\ \bibnamefont {Swanson}},\ }\href {\doibase 10.1016/j.ppnp.2016.11.003} {\bibfield  {journal} {\bibinfo  {journal} {Prog. Part. Nucl. Phys.}\ }\textbf {\bibinfo {volume} {93}},\ \bibinfo {pages} {143} (\bibinfo {year} {2017})},\ \Eprint {http://arxiv.org/abs/1610.04528} {arXiv:1610.04528 [hep-ph]} \BibitemShut {NoStop}%
\bibitem [{\citenamefont {Guo}\ \emph {et~al.}(2018)\citenamefont {Guo}, \citenamefont {Hanhart}, \citenamefont {Mei\ss{}ner}, \citenamefont {Wang}, \citenamefont {Zhao},\ and\ \citenamefont {Zou}}]{Guo:2017jvc}%
  \BibitemOpen
  \bibfield  {author} {\bibinfo {author} {\bibfnamefont {F.-K.}\ \bibnamefont {Guo}}, \bibinfo {author} {\bibfnamefont {C.}~\bibnamefont {Hanhart}}, \bibinfo {author} {\bibfnamefont {U.-G.}\ \bibnamefont {Mei\ss{}ner}}, \bibinfo {author} {\bibfnamefont {Q.}~\bibnamefont {Wang}}, \bibinfo {author} {\bibfnamefont {Q.}~\bibnamefont {Zhao}}, \ and\ \bibinfo {author} {\bibfnamefont {B.-S.}\ \bibnamefont {Zou}},\ }\href {\doibase 10.1103/RevModPhys.90.015004} {\bibfield  {journal} {\bibinfo  {journal} {Rev. Mod. Phys.}\ }\textbf {\bibinfo {volume} {90}},\ \bibinfo {pages} {015004} (\bibinfo {year} {2018})},\ \bibinfo {note} {[Erratum: Rev.Mod.Phys. 94, 029901 (2022)]},\ \Eprint {http://arxiv.org/abs/1705.00141} {arXiv:1705.00141 [hep-ph]} \BibitemShut {NoStop}%
\bibitem [{\citenamefont {Ali}\ \emph {et~al.}(2017)\citenamefont {Ali}, \citenamefont {Lange},\ and\ \citenamefont {Stone}}]{Ali:2017jda}%
  \BibitemOpen
  \bibfield  {author} {\bibinfo {author} {\bibfnamefont {A.}~\bibnamefont {Ali}}, \bibinfo {author} {\bibfnamefont {J.~S.}\ \bibnamefont {Lange}}, \ and\ \bibinfo {author} {\bibfnamefont {S.}~\bibnamefont {Stone}},\ }\href {\doibase 10.1016/j.ppnp.2017.08.003} {\bibfield  {journal} {\bibinfo  {journal} {Prog. Part. Nucl. Phys.}\ }\textbf {\bibinfo {volume} {97}},\ \bibinfo {pages} {123} (\bibinfo {year} {2017})},\ \Eprint {http://arxiv.org/abs/1706.00610} {arXiv:1706.00610 [hep-ph]} \BibitemShut {NoStop}%
\bibitem [{\citenamefont {Olsen}\ \emph {et~al.}(2018)\citenamefont {Olsen}, \citenamefont {Skwarnicki},\ and\ \citenamefont {Zieminska}}]{Olsen:2017bmm}%
  \BibitemOpen
  \bibfield  {author} {\bibinfo {author} {\bibfnamefont {S.~L.}\ \bibnamefont {Olsen}}, \bibinfo {author} {\bibfnamefont {T.}~\bibnamefont {Skwarnicki}}, \ and\ \bibinfo {author} {\bibfnamefont {D.}~\bibnamefont {Zieminska}},\ }\href {\doibase 10.1103/RevModPhys.90.015003} {\bibfield  {journal} {\bibinfo  {journal} {Rev. Mod. Phys.}\ }\textbf {\bibinfo {volume} {90}},\ \bibinfo {pages} {015003} (\bibinfo {year} {2018})},\ \Eprint {http://arxiv.org/abs/1708.04012} {arXiv:1708.04012 [hep-ph]} \BibitemShut {NoStop}%
\bibitem [{\citenamefont {Liu}\ \emph {et~al.}(2019)\citenamefont {Liu}, \citenamefont {Chen}, \citenamefont {Chen}, \citenamefont {Liu},\ and\ \citenamefont {Zhu}}]{Liu:2019zoy}%
  \BibitemOpen
  \bibfield  {author} {\bibinfo {author} {\bibfnamefont {Y.-R.}\ \bibnamefont {Liu}}, \bibinfo {author} {\bibfnamefont {H.-X.}\ \bibnamefont {Chen}}, \bibinfo {author} {\bibfnamefont {W.}~\bibnamefont {Chen}}, \bibinfo {author} {\bibfnamefont {X.}~\bibnamefont {Liu}}, \ and\ \bibinfo {author} {\bibfnamefont {S.-L.}\ \bibnamefont {Zhu}},\ }\href {\doibase 10.1016/j.ppnp.2019.04.003} {\bibfield  {journal} {\bibinfo  {journal} {Prog. Part. Nucl. Phys.}\ }\textbf {\bibinfo {volume} {107}},\ \bibinfo {pages} {237} (\bibinfo {year} {2019})},\ \Eprint {http://arxiv.org/abs/1903.11976} {arXiv:1903.11976 [hep-ph]} \BibitemShut {NoStop}%
\bibitem [{\citenamefont {Brambilla}\ \emph {et~al.}(2020)\citenamefont {Brambilla}, \citenamefont {Eidelman}, \citenamefont {Hanhart}, \citenamefont {Nefediev}, \citenamefont {Shen}, \citenamefont {Thomas}, \citenamefont {Vairo},\ and\ \citenamefont {Yuan}}]{Brambilla:2019esw}%
  \BibitemOpen
  \bibfield  {author} {\bibinfo {author} {\bibfnamefont {N.}~\bibnamefont {Brambilla}}, \bibinfo {author} {\bibfnamefont {S.}~\bibnamefont {Eidelman}}, \bibinfo {author} {\bibfnamefont {C.}~\bibnamefont {Hanhart}}, \bibinfo {author} {\bibfnamefont {A.}~\bibnamefont {Nefediev}}, \bibinfo {author} {\bibfnamefont {C.-P.}\ \bibnamefont {Shen}}, \bibinfo {author} {\bibfnamefont {C.~E.}\ \bibnamefont {Thomas}}, \bibinfo {author} {\bibfnamefont {A.}~\bibnamefont {Vairo}}, \ and\ \bibinfo {author} {\bibfnamefont {C.-Z.}\ \bibnamefont {Yuan}},\ }\href {\doibase 10.1016/j.physrep.2020.05.001} {\bibfield  {journal} {\bibinfo  {journal} {Phys. Rept.}\ }\textbf {\bibinfo {volume} {873}},\ \bibinfo {pages} {1} (\bibinfo {year} {2020})},\ \Eprint {http://arxiv.org/abs/1907.07583} {arXiv:1907.07583 [hep-ex]} \BibitemShut {NoStop}%
\bibitem [{\citenamefont {Chen}\ \emph {et~al.}(2023)\citenamefont {Chen}, \citenamefont {Chen}, \citenamefont {Liu}, \citenamefont {Liu},\ and\ \citenamefont {Zhu}}]{Chen:2022asf}%
  \BibitemOpen
  \bibfield  {author} {\bibinfo {author} {\bibfnamefont {H.-X.}\ \bibnamefont {Chen}}, \bibinfo {author} {\bibfnamefont {W.}~\bibnamefont {Chen}}, \bibinfo {author} {\bibfnamefont {X.}~\bibnamefont {Liu}}, \bibinfo {author} {\bibfnamefont {Y.-R.}\ \bibnamefont {Liu}}, \ and\ \bibinfo {author} {\bibfnamefont {S.-L.}\ \bibnamefont {Zhu}},\ }\href {\doibase 10.1088/1361-6633/aca3b6} {\bibfield  {journal} {\bibinfo  {journal} {Rept. Prog. Phys.}\ }\textbf {\bibinfo {volume} {86}},\ \bibinfo {pages} {026201} (\bibinfo {year} {2023})},\ \Eprint {http://arxiv.org/abs/2204.02649} {arXiv:2204.02649 [hep-ph]} \BibitemShut {NoStop}%
\bibitem [{\citenamefont {Luo}\ \emph {et~al.}(2022{\natexlab{a}})\citenamefont {Luo}, \citenamefont {Geng},\ and\ \citenamefont {Liu}}]{Luo:2022cun}%
  \BibitemOpen
  \bibfield  {author} {\bibinfo {author} {\bibfnamefont {S.-Q.}\ \bibnamefont {Luo}}, \bibinfo {author} {\bibfnamefont {L.-S.}\ \bibnamefont {Geng}}, \ and\ \bibinfo {author} {\bibfnamefont {X.}~\bibnamefont {Liu}},\ }\href {\doibase 10.1103/PhysRevD.106.014017} {\bibfield  {journal} {\bibinfo  {journal} {Phys. Rev. D}\ }\textbf {\bibinfo {volume} {106}},\ \bibinfo {pages} {014017} (\bibinfo {year} {2022}{\natexlab{a}})},\ \Eprint {http://arxiv.org/abs/2206.04586} {arXiv:2206.04586 [hep-ph]} \BibitemShut {NoStop}%
\bibitem [{\citenamefont {Hiyama}(2012)}]{Hiyama:2012sma}%
  \BibitemOpen
  \bibfield  {author} {\bibinfo {author} {\bibfnamefont {E.}~\bibnamefont {Hiyama}},\ }\href {\doibase 10.1093/ptep/pts015} {\bibfield  {journal} {\bibinfo  {journal} {PTEP}\ }\textbf {\bibinfo {volume} {2012}},\ \bibinfo {pages} {01A204} (\bibinfo {year} {2012})}\BibitemShut {NoStop}%
\bibitem [{\citenamefont {Hiyama}\ \emph {et~al.}(2003)\citenamefont {Hiyama}, \citenamefont {Kino},\ and\ \citenamefont {Kamimura}}]{Hiyama:2003cu}%
  \BibitemOpen
  \bibfield  {author} {\bibinfo {author} {\bibfnamefont {E.}~\bibnamefont {Hiyama}}, \bibinfo {author} {\bibfnamefont {Y.}~\bibnamefont {Kino}}, \ and\ \bibinfo {author} {\bibfnamefont {M.}~\bibnamefont {Kamimura}},\ }\href {\doibase 10.1016/S0146-6410(03)90015-9} {\bibfield  {journal} {\bibinfo  {journal} {Prog. Part. Nucl. Phys.}\ }\textbf {\bibinfo {volume} {51}},\ \bibinfo {pages} {223} (\bibinfo {year} {2003})}\BibitemShut {NoStop}%
\bibitem [{\citenamefont {Montesinos}\ \emph {et~al.}(2024)\citenamefont {Montesinos}, \citenamefont {Song}, \citenamefont {Liang}, \citenamefont {Oset}, \citenamefont {Nieves},\ and\ \citenamefont {Albaladejo}}]{Montesinos:2024eoy}%
  \BibitemOpen
  \bibfield  {author} {\bibinfo {author} {\bibfnamefont {V.}~\bibnamefont {Montesinos}}, \bibinfo {author} {\bibfnamefont {J.}~\bibnamefont {Song}}, \bibinfo {author} {\bibfnamefont {W.-H.}\ \bibnamefont {Liang}}, \bibinfo {author} {\bibfnamefont {E.}~\bibnamefont {Oset}}, \bibinfo {author} {\bibfnamefont {J.}~\bibnamefont {Nieves}}, \ and\ \bibinfo {author} {\bibfnamefont {M.}~\bibnamefont {Albaladejo}},\ }\href {\doibase 10.1103/PhysRevD.110.054043} {\bibfield  {journal} {\bibinfo  {journal} {Phys. Rev. D}\ }\textbf {\bibinfo {volume} {110}},\ \bibinfo {pages} {054043} (\bibinfo {year} {2024})},\ \Eprint {http://arxiv.org/abs/2405.09467} {arXiv:2405.09467 [hep-ph]} \BibitemShut {NoStop}%
\bibitem [{\citenamefont {Luo}\ \emph {et~al.}(2022{\natexlab{b}})\citenamefont {Luo}, \citenamefont {Wu}, \citenamefont {Liu}, \citenamefont {Geng},\ and\ \citenamefont {Liu}}]{Luo:2021ggs}%
  \BibitemOpen
  \bibfield  {author} {\bibinfo {author} {\bibfnamefont {S.-Q.}\ \bibnamefont {Luo}}, \bibinfo {author} {\bibfnamefont {T.-W.}\ \bibnamefont {Wu}}, \bibinfo {author} {\bibfnamefont {M.-Z.}\ \bibnamefont {Liu}}, \bibinfo {author} {\bibfnamefont {L.-S.}\ \bibnamefont {Geng}}, \ and\ \bibinfo {author} {\bibfnamefont {X.}~\bibnamefont {Liu}},\ }\href {\doibase 10.1103/PhysRevD.105.074033} {\bibfield  {journal} {\bibinfo  {journal} {Phys. Rev. D}\ }\textbf {\bibinfo {volume} {105}},\ \bibinfo {pages} {074033} (\bibinfo {year} {2022}{\natexlab{b}})},\ \Eprint {http://arxiv.org/abs/2111.15079} {arXiv:2111.15079 [hep-ph]} \BibitemShut {NoStop}%
\bibitem [{\citenamefont {Bayar}\ \emph {et~al.}(2023)\citenamefont {Bayar}, \citenamefont {Martinez~Torres}, \citenamefont {Khemchandani}, \citenamefont {Molina},\ and\ \citenamefont {Oset}}]{Bayar:2022bnc}%
  \BibitemOpen
  \bibfield  {author} {\bibinfo {author} {\bibfnamefont {M.}~\bibnamefont {Bayar}}, \bibinfo {author} {\bibfnamefont {A.}~\bibnamefont {Martinez~Torres}}, \bibinfo {author} {\bibfnamefont {K.~P.}\ \bibnamefont {Khemchandani}}, \bibinfo {author} {\bibfnamefont {R.}~\bibnamefont {Molina}}, \ and\ \bibinfo {author} {\bibfnamefont {E.}~\bibnamefont {Oset}},\ }\href {\doibase 10.1140/epjc/s10052-023-11207-5} {\bibfield  {journal} {\bibinfo  {journal} {Eur. Phys. J. C}\ }\textbf {\bibinfo {volume} {83}},\ \bibinfo {pages} {46} (\bibinfo {year} {2023})},\ \Eprint {http://arxiv.org/abs/2211.09294} {arXiv:2211.09294 [hep-ph]} \BibitemShut {NoStop}%
\bibitem [{\citenamefont {Ortega}(2024)}]{Ortega:2024ecy}%
  \BibitemOpen
  \bibfield  {author} {\bibinfo {author} {\bibfnamefont {P.~G.}\ \bibnamefont {Ortega}},\ }\href {\doibase 10.1103/PhysRevD.110.034015} {\bibfield  {journal} {\bibinfo  {journal} {Phys. Rev. D}\ }\textbf {\bibinfo {volume} {110}},\ \bibinfo {pages} {034015} (\bibinfo {year} {2024})},\ \Eprint {http://arxiv.org/abs/2403.10244} {arXiv:2403.10244 [hep-ph]} \BibitemShut {NoStop}%
\bibitem [{\citenamefont {Ikeno}\ \emph {et~al.}(2023)\citenamefont {Ikeno}, \citenamefont {Bayar},\ and\ \citenamefont {Oset}}]{Ikeno:2022jbb}%
  \BibitemOpen
  \bibfield  {author} {\bibinfo {author} {\bibfnamefont {N.}~\bibnamefont {Ikeno}}, \bibinfo {author} {\bibfnamefont {M.}~\bibnamefont {Bayar}}, \ and\ \bibinfo {author} {\bibfnamefont {E.}~\bibnamefont {Oset}},\ }\href {\doibase 10.1103/PhysRevD.107.034006} {\bibfield  {journal} {\bibinfo  {journal} {Phys. Rev. D}\ }\textbf {\bibinfo {volume} {107}},\ \bibinfo {pages} {034006} (\bibinfo {year} {2023})},\ \Eprint {http://arxiv.org/abs/2208.03698} {arXiv:2208.03698 [hep-ph]} \BibitemShut {NoStop}%
\bibitem [{\citenamefont {Debastiani}\ \emph {et~al.}(2017)\citenamefont {Debastiani}, \citenamefont {Dias},\ and\ \citenamefont {Oset}}]{Debastiani:2017vhv}%
  \BibitemOpen
  \bibfield  {author} {\bibinfo {author} {\bibfnamefont {V.~R.}\ \bibnamefont {Debastiani}}, \bibinfo {author} {\bibfnamefont {J.~M.}\ \bibnamefont {Dias}}, \ and\ \bibinfo {author} {\bibfnamefont {E.}~\bibnamefont {Oset}},\ }\href {\doibase 10.1103/PhysRevD.96.016014} {\bibfield  {journal} {\bibinfo  {journal} {Phys. Rev. D}\ }\textbf {\bibinfo {volume} {96}},\ \bibinfo {pages} {016014} (\bibinfo {year} {2017})},\ \Eprint {http://arxiv.org/abs/1705.09257} {arXiv:1705.09257 [hep-ph]} \BibitemShut {NoStop}%
\bibitem [{\citenamefont {Wu}\ \emph {et~al.}(2021{\natexlab{a}})\citenamefont {Wu}, \citenamefont {Liu},\ and\ \citenamefont {Geng}}]{Wu:2021dwy}%
  \BibitemOpen
  \bibfield  {author} {\bibinfo {author} {\bibfnamefont {T.-W.}\ \bibnamefont {Wu}}, \bibinfo {author} {\bibfnamefont {M.-Z.}\ \bibnamefont {Liu}}, \ and\ \bibinfo {author} {\bibfnamefont {L.-S.}\ \bibnamefont {Geng}},\ }\href {\doibase 10.1007/s00601-021-01619-y} {\bibfield  {journal} {\bibinfo  {journal} {Few Body Syst.}\ }\textbf {\bibinfo {volume} {62}},\ \bibinfo {pages} {38} (\bibinfo {year} {2021}{\natexlab{a}})},\ \Eprint {http://arxiv.org/abs/2105.09017} {arXiv:2105.09017 [hep-ph]} \BibitemShut {NoStop}%
\bibitem [{\citenamefont {Wu}\ \emph {et~al.}(2021{\natexlab{b}})\citenamefont {Wu}, \citenamefont {Liu},\ and\ \citenamefont {Geng}}]{Wu:2020job}%
  \BibitemOpen
  \bibfield  {author} {\bibinfo {author} {\bibfnamefont {T.-W.}\ \bibnamefont {Wu}}, \bibinfo {author} {\bibfnamefont {M.-Z.}\ \bibnamefont {Liu}}, \ and\ \bibinfo {author} {\bibfnamefont {L.-S.}\ \bibnamefont {Geng}},\ }\href {\doibase 10.1103/PhysRevD.103.L031501} {\bibfield  {journal} {\bibinfo  {journal} {Phys. Rev. D}\ }\textbf {\bibinfo {volume} {103}},\ \bibinfo {pages} {L031501} (\bibinfo {year} {2021}{\natexlab{b}})},\ \Eprint {http://arxiv.org/abs/2012.01134} {arXiv:2012.01134 [hep-ph]} \BibitemShut {NoStop}%
\bibitem [{\citenamefont {Ren}\ \emph {et~al.}(2019)\citenamefont {Ren}, \citenamefont {Malabarba}, \citenamefont {Geng}, \citenamefont {Khemchandani},\ and\ \citenamefont {Torres}}]{Ren:2019bne}%
  \BibitemOpen
  \bibfield  {author} {\bibinfo {author} {\bibfnamefont {X.-L.}\ \bibnamefont {Ren}}, \bibinfo {author} {\bibfnamefont {B.~B.}\ \bibnamefont {Malabarba}}, \bibinfo {author} {\bibfnamefont {L.-S.}\ \bibnamefont {Geng}}, \bibinfo {author} {\bibfnamefont {K.~P.}\ \bibnamefont {Khemchandani}}, \ and\ \bibinfo {author} {\bibfnamefont {A.~M.}\ \bibnamefont {Torres}},\ }\href {\doibase 10.7566/JPSCP.26.031006} {\bibfield  {journal} {\bibinfo  {journal} {JPS Conf. Proc.}\ }\textbf {\bibinfo {volume} {26}},\ \bibinfo {pages} {031006} (\bibinfo {year} {2019})}\BibitemShut {NoStop}%
\bibitem [{\citenamefont {Ren}\ \emph {et~al.}(2024)\citenamefont {Ren}, \citenamefont {Khemchandani},\ and\ \citenamefont {Mart\'\i{}nez~Torres}}]{Ren:2024mjh}%
  \BibitemOpen
  \bibfield  {author} {\bibinfo {author} {\bibfnamefont {X.-L.}\ \bibnamefont {Ren}}, \bibinfo {author} {\bibfnamefont {K.~P.}\ \bibnamefont {Khemchandani}}, \ and\ \bibinfo {author} {\bibfnamefont {A.}~\bibnamefont {Mart\'\i{}nez~Torres}},\ }\href@noop {} {\  (\bibinfo {year} {2024})},\ \Eprint {http://arxiv.org/abs/2409.16281} {arXiv:2409.16281 [hep-ph]} \BibitemShut {NoStop}%
\bibitem [{\citenamefont {Pang}\ \emph {et~al.}(2020)\citenamefont {Pang}, \citenamefont {Wu},\ and\ \citenamefont {Geng}}]{Pang:2020pkl}%
  \BibitemOpen
  \bibfield  {author} {\bibinfo {author} {\bibfnamefont {J.-Y.}\ \bibnamefont {Pang}}, \bibinfo {author} {\bibfnamefont {J.-J.}\ \bibnamefont {Wu}}, \ and\ \bibinfo {author} {\bibfnamefont {L.-S.}\ \bibnamefont {Geng}},\ }\href {\doibase 10.1103/PhysRevD.102.114515} {\bibfield  {journal} {\bibinfo  {journal} {Phys. Rev. D}\ }\textbf {\bibinfo {volume} {102}},\ \bibinfo {pages} {114515} (\bibinfo {year} {2020})},\ \Eprint {http://arxiv.org/abs/2008.13014} {arXiv:2008.13014 [hep-lat]} \BibitemShut {NoStop}%
\bibitem [{\citenamefont {Wu}\ \emph {et~al.}(2019)\citenamefont {Wu}, \citenamefont {Liu}, \citenamefont {Geng}, \citenamefont {Hiyama},\ and\ \citenamefont {Valderrama}}]{Wu:2019vsy}%
  \BibitemOpen
  \bibfield  {author} {\bibinfo {author} {\bibfnamefont {T.-W.}\ \bibnamefont {Wu}}, \bibinfo {author} {\bibfnamefont {M.-Z.}\ \bibnamefont {Liu}}, \bibinfo {author} {\bibfnamefont {L.-S.}\ \bibnamefont {Geng}}, \bibinfo {author} {\bibfnamefont {E.}~\bibnamefont {Hiyama}}, \ and\ \bibinfo {author} {\bibfnamefont {M.~P.}\ \bibnamefont {Valderrama}},\ }\href {\doibase 10.1103/PhysRevD.100.034029} {\bibfield  {journal} {\bibinfo  {journal} {Phys. Rev. D}\ }\textbf {\bibinfo {volume} {100}},\ \bibinfo {pages} {034029} (\bibinfo {year} {2019})},\ \Eprint {http://arxiv.org/abs/1906.11995} {arXiv:1906.11995 [hep-ph]} \BibitemShut {NoStop}%
\bibitem [{\citenamefont {Martinez~Torres}\ \emph {et~al.}(2019)\citenamefont {Martinez~Torres}, \citenamefont {Khemchandani},\ and\ \citenamefont {Geng}}]{MartinezTorres:2018zbl}%
  \BibitemOpen
  \bibfield  {author} {\bibinfo {author} {\bibfnamefont {A.}~\bibnamefont {Martinez~Torres}}, \bibinfo {author} {\bibfnamefont {K.~P.}\ \bibnamefont {Khemchandani}}, \ and\ \bibinfo {author} {\bibfnamefont {L.-S.}\ \bibnamefont {Geng}},\ }\href {\doibase 10.1103/PhysRevD.99.076017} {\bibfield  {journal} {\bibinfo  {journal} {Phys. Rev. D}\ }\textbf {\bibinfo {volume} {99}},\ \bibinfo {pages} {076017} (\bibinfo {year} {2019})},\ \Eprint {http://arxiv.org/abs/1809.01059} {arXiv:1809.01059 [hep-ph]} \BibitemShut {NoStop}%
\bibitem [{\citenamefont {van Beveren}\ and\ \citenamefont {Rupp}(2003)}]{vanBeveren:2003kd}%
  \BibitemOpen
  \bibfield  {author} {\bibinfo {author} {\bibfnamefont {E.}~\bibnamefont {van Beveren}}\ and\ \bibinfo {author} {\bibfnamefont {G.}~\bibnamefont {Rupp}},\ }\href {\doibase 10.1103/PhysRevLett.91.012003} {\bibfield  {journal} {\bibinfo  {journal} {Phys. Rev. Lett.}\ }\textbf {\bibinfo {volume} {91}},\ \bibinfo {pages} {012003} (\bibinfo {year} {2003})},\ \Eprint {http://arxiv.org/abs/hep-ph/0305035} {arXiv:hep-ph/0305035} \BibitemShut {NoStop}%
\bibitem [{\citenamefont {Barnes}\ \emph {et~al.}(2003)\citenamefont {Barnes}, \citenamefont {Close},\ and\ \citenamefont {Lipkin}}]{Barnes:2003dj}%
  \BibitemOpen
  \bibfield  {author} {\bibinfo {author} {\bibfnamefont {T.}~\bibnamefont {Barnes}}, \bibinfo {author} {\bibfnamefont {F.~E.}\ \bibnamefont {Close}}, \ and\ \bibinfo {author} {\bibfnamefont {H.~J.}\ \bibnamefont {Lipkin}},\ }\href {\doibase 10.1103/PhysRevD.68.054006} {\bibfield  {journal} {\bibinfo  {journal} {Phys. Rev. D}\ }\textbf {\bibinfo {volume} {68}},\ \bibinfo {pages} {054006} (\bibinfo {year} {2003})},\ \Eprint {http://arxiv.org/abs/hep-ph/0305025} {arXiv:hep-ph/0305025} \BibitemShut {NoStop}%
\bibitem [{\citenamefont {Chen}\ and\ \citenamefont {Li}(2004)}]{Chen:2004dy}%
  \BibitemOpen
  \bibfield  {author} {\bibinfo {author} {\bibfnamefont {Y.-Q.}\ \bibnamefont {Chen}}\ and\ \bibinfo {author} {\bibfnamefont {X.-Q.}\ \bibnamefont {Li}},\ }\href {\doibase 10.1103/PhysRevLett.93.232001} {\bibfield  {journal} {\bibinfo  {journal} {Phys. Rev. Lett.}\ }\textbf {\bibinfo {volume} {93}},\ \bibinfo {pages} {232001} (\bibinfo {year} {2004})},\ \Eprint {http://arxiv.org/abs/hep-ph/0407062} {arXiv:hep-ph/0407062} \BibitemShut {NoStop}%
\bibitem [{\citenamefont {Kolomeitsev}\ and\ \citenamefont {Lutz}(2004)}]{Kolomeitsev:2003ac}%
  \BibitemOpen
  \bibfield  {author} {\bibinfo {author} {\bibfnamefont {E.~E.}\ \bibnamefont {Kolomeitsev}}\ and\ \bibinfo {author} {\bibfnamefont {M.~F.~M.}\ \bibnamefont {Lutz}},\ }\href {\doibase 10.1016/j.physletb.2003.10.118} {\bibfield  {journal} {\bibinfo  {journal} {Phys. Lett. B}\ }\textbf {\bibinfo {volume} {582}},\ \bibinfo {pages} {39} (\bibinfo {year} {2004})},\ \Eprint {http://arxiv.org/abs/hep-ph/0307133} {arXiv:hep-ph/0307133} \BibitemShut {NoStop}%
\bibitem [{\citenamefont {Gamermann}\ \emph {et~al.}(2007)\citenamefont {Gamermann}, \citenamefont {Oset}, \citenamefont {Strottman},\ and\ \citenamefont {Vicente~Vacas}}]{Gamermann:2006nm}%
  \BibitemOpen
  \bibfield  {author} {\bibinfo {author} {\bibfnamefont {D.}~\bibnamefont {Gamermann}}, \bibinfo {author} {\bibfnamefont {E.}~\bibnamefont {Oset}}, \bibinfo {author} {\bibfnamefont {D.}~\bibnamefont {Strottman}}, \ and\ \bibinfo {author} {\bibfnamefont {M.~J.}\ \bibnamefont {Vicente~Vacas}},\ }\href {\doibase 10.1103/PhysRevD.76.074016} {\bibfield  {journal} {\bibinfo  {journal} {Phys. Rev. D}\ }\textbf {\bibinfo {volume} {76}},\ \bibinfo {pages} {074016} (\bibinfo {year} {2007})},\ \Eprint {http://arxiv.org/abs/hep-ph/0612179} {arXiv:hep-ph/0612179} \BibitemShut {NoStop}%
\bibitem [{\citenamefont {Guo}\ \emph {et~al.}(2007)\citenamefont {Guo}, \citenamefont {Shen},\ and\ \citenamefont {Chiang}}]{Guo:2006rp}%
  \BibitemOpen
  \bibfield  {author} {\bibinfo {author} {\bibfnamefont {F.-K.}\ \bibnamefont {Guo}}, \bibinfo {author} {\bibfnamefont {P.-N.}\ \bibnamefont {Shen}}, \ and\ \bibinfo {author} {\bibfnamefont {H.-C.}\ \bibnamefont {Chiang}},\ }\href {\doibase 10.1016/j.physletb.2007.01.050} {\bibfield  {journal} {\bibinfo  {journal} {Phys. Lett. B}\ }\textbf {\bibinfo {volume} {647}},\ \bibinfo {pages} {133} (\bibinfo {year} {2007})},\ \Eprint {http://arxiv.org/abs/hep-ph/0610008} {arXiv:hep-ph/0610008} \BibitemShut {NoStop}%
\bibitem [{\citenamefont {Yang}\ \emph {et~al.}(2022)\citenamefont {Yang}, \citenamefont {Wang}, \citenamefont {Wu}, \citenamefont {Oka},\ and\ \citenamefont {Zhu}}]{Yang:2021tvc}%
  \BibitemOpen
  \bibfield  {author} {\bibinfo {author} {\bibfnamefont {Z.}~\bibnamefont {Yang}}, \bibinfo {author} {\bibfnamefont {G.-J.}\ \bibnamefont {Wang}}, \bibinfo {author} {\bibfnamefont {J.-J.}\ \bibnamefont {Wu}}, \bibinfo {author} {\bibfnamefont {M.}~\bibnamefont {Oka}}, \ and\ \bibinfo {author} {\bibfnamefont {S.-L.}\ \bibnamefont {Zhu}},\ }\href {\doibase 10.1103/PhysRevLett.128.112001} {\bibfield  {journal} {\bibinfo  {journal} {Phys. Rev. Lett.}\ }\textbf {\bibinfo {volume} {128}},\ \bibinfo {pages} {112001} (\bibinfo {year} {2022})},\ \Eprint {http://arxiv.org/abs/2107.04860} {arXiv:2107.04860 [hep-ph]} \BibitemShut {NoStop}%
\bibitem [{\citenamefont {Liu}\ \emph {et~al.}(2022)\citenamefont {Liu}, \citenamefont {Ling}, \citenamefont {Geng}, \citenamefont {En-Wang},\ and\ \citenamefont {Xie}}]{Liu:2022dmm}%
  \BibitemOpen
  \bibfield  {author} {\bibinfo {author} {\bibfnamefont {M.-Z.}\ \bibnamefont {Liu}}, \bibinfo {author} {\bibfnamefont {X.-Z.}\ \bibnamefont {Ling}}, \bibinfo {author} {\bibfnamefont {L.-S.}\ \bibnamefont {Geng}}, \bibinfo {author} {\bibnamefont {En-Wang}}, \ and\ \bibinfo {author} {\bibfnamefont {J.-J.}\ \bibnamefont {Xie}},\ }\href {\doibase 10.1103/PhysRevD.106.114011} {\bibfield  {journal} {\bibinfo  {journal} {Phys. Rev. D}\ }\textbf {\bibinfo {volume} {106}},\ \bibinfo {pages} {114011} (\bibinfo {year} {2022})},\ \Eprint {http://arxiv.org/abs/2209.01103} {arXiv:2209.01103 [hep-ph]} \BibitemShut {NoStop}%
\bibitem [{\citenamefont {Mohler}\ \emph {et~al.}(2013)\citenamefont {Mohler}, \citenamefont {Lang}, \citenamefont {Leskovec}, \citenamefont {Prelovsek},\ and\ \citenamefont {Woloshyn}}]{Mohler:2013rwa}%
  \BibitemOpen
  \bibfield  {author} {\bibinfo {author} {\bibfnamefont {D.}~\bibnamefont {Mohler}}, \bibinfo {author} {\bibfnamefont {C.~B.}\ \bibnamefont {Lang}}, \bibinfo {author} {\bibfnamefont {L.}~\bibnamefont {Leskovec}}, \bibinfo {author} {\bibfnamefont {S.}~\bibnamefont {Prelovsek}}, \ and\ \bibinfo {author} {\bibfnamefont {R.~M.}\ \bibnamefont {Woloshyn}},\ }\href {\doibase 10.1103/PhysRevLett.111.222001} {\bibfield  {journal} {\bibinfo  {journal} {Phys. Rev. Lett.}\ }\textbf {\bibinfo {volume} {111}},\ \bibinfo {pages} {222001} (\bibinfo {year} {2013})},\ \Eprint {http://arxiv.org/abs/1308.3175} {arXiv:1308.3175 [hep-lat]} \BibitemShut {NoStop}%
\bibitem [{\citenamefont {Lang}\ \emph {et~al.}(2014)\citenamefont {Lang}, \citenamefont {Leskovec}, \citenamefont {Mohler}, \citenamefont {Prelovsek},\ and\ \citenamefont {Woloshyn}}]{Lang:2014yfa}%
  \BibitemOpen
  \bibfield  {author} {\bibinfo {author} {\bibfnamefont {C.~B.}\ \bibnamefont {Lang}}, \bibinfo {author} {\bibfnamefont {L.}~\bibnamefont {Leskovec}}, \bibinfo {author} {\bibfnamefont {D.}~\bibnamefont {Mohler}}, \bibinfo {author} {\bibfnamefont {S.}~\bibnamefont {Prelovsek}}, \ and\ \bibinfo {author} {\bibfnamefont {R.~M.}\ \bibnamefont {Woloshyn}},\ }\href {\doibase 10.1103/PhysRevD.90.034510} {\bibfield  {journal} {\bibinfo  {journal} {Phys. Rev. D}\ }\textbf {\bibinfo {volume} {90}},\ \bibinfo {pages} {034510} (\bibinfo {year} {2014})},\ \Eprint {http://arxiv.org/abs/1403.8103} {arXiv:1403.8103 [hep-lat]} \BibitemShut {NoStop}%
\bibitem [{\citenamefont {Bali}\ \emph {et~al.}(2017)\citenamefont {Bali}, \citenamefont {Collins}, \citenamefont {Cox},\ and\ \citenamefont {Sch\"afer}}]{Bali:2017pdv}%
  \BibitemOpen
  \bibfield  {author} {\bibinfo {author} {\bibfnamefont {G.~S.}\ \bibnamefont {Bali}}, \bibinfo {author} {\bibfnamefont {S.}~\bibnamefont {Collins}}, \bibinfo {author} {\bibfnamefont {A.}~\bibnamefont {Cox}}, \ and\ \bibinfo {author} {\bibfnamefont {A.}~\bibnamefont {Sch\"afer}},\ }\href {\doibase 10.1103/PhysRevD.96.074501} {\bibfield  {journal} {\bibinfo  {journal} {Phys. Rev. D}\ }\textbf {\bibinfo {volume} {96}},\ \bibinfo {pages} {074501} (\bibinfo {year} {2017})},\ \Eprint {http://arxiv.org/abs/1706.01247} {arXiv:1706.01247 [hep-lat]} \BibitemShut {NoStop}%
\bibitem [{\citenamefont {Cheung}\ \emph {et~al.}(2021)\citenamefont {Cheung}, \citenamefont {Thomas}, \citenamefont {Wilson}, \citenamefont {Moir}, \citenamefont {Peardon},\ and\ \citenamefont {Ryan}}]{Cheung:2020mql}%
  \BibitemOpen
  \bibfield  {author} {\bibinfo {author} {\bibfnamefont {G.~K.~C.}\ \bibnamefont {Cheung}}, \bibinfo {author} {\bibfnamefont {C.~E.}\ \bibnamefont {Thomas}}, \bibinfo {author} {\bibfnamefont {D.~J.}\ \bibnamefont {Wilson}}, \bibinfo {author} {\bibfnamefont {G.}~\bibnamefont {Moir}}, \bibinfo {author} {\bibfnamefont {M.}~\bibnamefont {Peardon}}, \ and\ \bibinfo {author} {\bibfnamefont {S.~M.}\ \bibnamefont {Ryan}} (\bibinfo {collaboration} {Hadron Spectrum}),\ }\href {\doibase 10.1007/JHEP02(2021)100} {\bibfield  {journal} {\bibinfo  {journal} {JHEP}\ }\textbf {\bibinfo {volume} {02}},\ \bibinfo {pages} {100} (\bibinfo {year} {2021})},\ \Eprint {http://arxiv.org/abs/2008.06432} {arXiv:2008.06432 [hep-lat]} \BibitemShut {NoStop}%
\bibitem [{\citenamefont {Mart\'\i{}nez~Torres}\ \emph {et~al.}(2015)\citenamefont {Mart\'\i{}nez~Torres}, \citenamefont {Oset}, \citenamefont {Prelovsek},\ and\ \citenamefont {Ramos}}]{MartinezTorres:2014kpc}%
  \BibitemOpen
  \bibfield  {author} {\bibinfo {author} {\bibfnamefont {A.}~\bibnamefont {Mart\'\i{}nez~Torres}}, \bibinfo {author} {\bibfnamefont {E.}~\bibnamefont {Oset}}, \bibinfo {author} {\bibfnamefont {S.}~\bibnamefont {Prelovsek}}, \ and\ \bibinfo {author} {\bibfnamefont {A.}~\bibnamefont {Ramos}},\ }\href {\doibase 10.1007/JHEP05(2015)153} {\bibfield  {journal} {\bibinfo  {journal} {JHEP}\ }\textbf {\bibinfo {volume} {05}},\ \bibinfo {pages} {153} (\bibinfo {year} {2015})},\ \Eprint {http://arxiv.org/abs/1412.1706} {arXiv:1412.1706 [hep-lat]} \BibitemShut {NoStop}%
\bibitem [{\citenamefont {Molina}\ \emph {et~al.}(2010)\citenamefont {Molina}, \citenamefont {Branz},\ and\ \citenamefont {Oset}}]{Molina:2010tx}%
  \BibitemOpen
  \bibfield  {author} {\bibinfo {author} {\bibfnamefont {R.}~\bibnamefont {Molina}}, \bibinfo {author} {\bibfnamefont {T.}~\bibnamefont {Branz}}, \ and\ \bibinfo {author} {\bibfnamefont {E.}~\bibnamefont {Oset}},\ }\href {\doibase 10.1103/PhysRevD.82.014010} {\bibfield  {journal} {\bibinfo  {journal} {Phys. Rev. D}\ }\textbf {\bibinfo {volume} {82}},\ \bibinfo {pages} {014010} (\bibinfo {year} {2010})},\ \Eprint {http://arxiv.org/abs/1005.0335} {arXiv:1005.0335 [hep-ph]} \BibitemShut {NoStop}%
\bibitem [{\citenamefont {Aaij}\ \emph {et~al.}(2020{\natexlab{a}})\citenamefont {Aaij} \emph {et~al.}}]{LHCb:2020bls}%
  \BibitemOpen
  \bibfield  {author} {\bibinfo {author} {\bibfnamefont {R.}~\bibnamefont {Aaij}} \emph {et~al.} (\bibinfo {collaboration} {LHCb}),\ }\href {\doibase 10.1103/PhysRevLett.125.242001} {\bibfield  {journal} {\bibinfo  {journal} {Phys. Rev. Lett.}\ }\textbf {\bibinfo {volume} {125}},\ \bibinfo {pages} {242001} (\bibinfo {year} {2020}{\natexlab{a}})},\ \Eprint {http://arxiv.org/abs/2009.00025} {arXiv:2009.00025 [hep-ex]} \BibitemShut {NoStop}%
\bibitem [{\citenamefont {Aaij}\ \emph {et~al.}(2020{\natexlab{b}})\citenamefont {Aaij} \emph {et~al.}}]{LHCb:2020pxc}%
  \BibitemOpen
  \bibfield  {author} {\bibinfo {author} {\bibfnamefont {R.}~\bibnamefont {Aaij}} \emph {et~al.} (\bibinfo {collaboration} {LHCb}),\ }\href {\doibase 10.1103/PhysRevD.102.112003} {\bibfield  {journal} {\bibinfo  {journal} {Phys. Rev. D}\ }\textbf {\bibinfo {volume} {102}},\ \bibinfo {pages} {112003} (\bibinfo {year} {2020}{\natexlab{b}})},\ \Eprint {http://arxiv.org/abs/2009.00026} {arXiv:2009.00026 [hep-ex]} \BibitemShut {NoStop}%
\bibitem [{\citenamefont {Molina}\ and\ \citenamefont {Oset}(2020)}]{Molina:2020hde}%
  \BibitemOpen
  \bibfield  {author} {\bibinfo {author} {\bibfnamefont {R.}~\bibnamefont {Molina}}\ and\ \bibinfo {author} {\bibfnamefont {E.}~\bibnamefont {Oset}},\ }\href {\doibase 10.1016/j.physletb.2020.135870} {\bibfield  {journal} {\bibinfo  {journal} {Phys. Lett. B}\ }\textbf {\bibinfo {volume} {811}},\ \bibinfo {pages} {135870} (\bibinfo {year} {2020})},\ \bibinfo {note} {[Erratum: Phys.Lett.B 837, 137645 (2023)]},\ \Eprint {http://arxiv.org/abs/2008.11171} {arXiv:2008.11171 [hep-ph]} \BibitemShut {NoStop}%
\bibitem [{\citenamefont {Liu}\ \emph {et~al.}(2020)\citenamefont {Liu}, \citenamefont {Xie},\ and\ \citenamefont {Geng}}]{Liu:2020nil}%
  \BibitemOpen
  \bibfield  {author} {\bibinfo {author} {\bibfnamefont {M.-Z.}\ \bibnamefont {Liu}}, \bibinfo {author} {\bibfnamefont {J.-J.}\ \bibnamefont {Xie}}, \ and\ \bibinfo {author} {\bibfnamefont {L.-S.}\ \bibnamefont {Geng}},\ }\href {\doibase 10.1103/PhysRevD.102.091502} {\bibfield  {journal} {\bibinfo  {journal} {Phys. Rev. D}\ }\textbf {\bibinfo {volume} {102}},\ \bibinfo {pages} {091502} (\bibinfo {year} {2020})},\ \Eprint {http://arxiv.org/abs/2008.07389} {arXiv:2008.07389 [hep-ph]} \BibitemShut {NoStop}%
\bibitem [{\citenamefont {Huang}\ \emph {et~al.}(2020)\citenamefont {Huang}, \citenamefont {Lu}, \citenamefont {Xie},\ and\ \citenamefont {Geng}}]{Huang:2020ptc}%
  \BibitemOpen
  \bibfield  {author} {\bibinfo {author} {\bibfnamefont {Y.}~\bibnamefont {Huang}}, \bibinfo {author} {\bibfnamefont {J.-X.}\ \bibnamefont {Lu}}, \bibinfo {author} {\bibfnamefont {J.-J.}\ \bibnamefont {Xie}}, \ and\ \bibinfo {author} {\bibfnamefont {L.-S.}\ \bibnamefont {Geng}},\ }\href {\doibase 10.1140/epjc/s10052-020-08516-4} {\bibfield  {journal} {\bibinfo  {journal} {Eur. Phys. J. C}\ }\textbf {\bibinfo {volume} {80}},\ \bibinfo {pages} {973} (\bibinfo {year} {2020})},\ \Eprint {http://arxiv.org/abs/2008.07959} {arXiv:2008.07959 [hep-ph]} \BibitemShut {NoStop}%
\bibitem [{\citenamefont {Hu}\ \emph {et~al.}(2021)\citenamefont {Hu}, \citenamefont {Lao}, \citenamefont {Ling},\ and\ \citenamefont {Wang}}]{Hu:2020mxp}%
  \BibitemOpen
  \bibfield  {author} {\bibinfo {author} {\bibfnamefont {M.-W.}\ \bibnamefont {Hu}}, \bibinfo {author} {\bibfnamefont {X.-Y.}\ \bibnamefont {Lao}}, \bibinfo {author} {\bibfnamefont {P.}~\bibnamefont {Ling}}, \ and\ \bibinfo {author} {\bibfnamefont {Q.}~\bibnamefont {Wang}},\ }\href {\doibase 10.1088/1674-1137/abcfaa} {\bibfield  {journal} {\bibinfo  {journal} {Chin. Phys. C}\ }\textbf {\bibinfo {volume} {45}},\ \bibinfo {pages} {021003} (\bibinfo {year} {2021})},\ \Eprint {http://arxiv.org/abs/2008.06894} {arXiv:2008.06894 [hep-ph]} \BibitemShut {NoStop}%
\bibitem [{\citenamefont {Chen}\ \emph {et~al.}(2020)\citenamefont {Chen}, \citenamefont {Chen}, \citenamefont {Dong},\ and\ \citenamefont {Su}}]{Chen:2020aos}%
  \BibitemOpen
  \bibfield  {author} {\bibinfo {author} {\bibfnamefont {H.-X.}\ \bibnamefont {Chen}}, \bibinfo {author} {\bibfnamefont {W.}~\bibnamefont {Chen}}, \bibinfo {author} {\bibfnamefont {R.-R.}\ \bibnamefont {Dong}}, \ and\ \bibinfo {author} {\bibfnamefont {N.}~\bibnamefont {Su}},\ }\href {\doibase 10.1088/0256-307X/37/10/101201} {\bibfield  {journal} {\bibinfo  {journal} {Chin. Phys. Lett.}\ }\textbf {\bibinfo {volume} {37}},\ \bibinfo {pages} {101201} (\bibinfo {year} {2020})},\ \Eprint {http://arxiv.org/abs/2008.07516} {arXiv:2008.07516 [hep-ph]} \BibitemShut {NoStop}%
\bibitem [{\citenamefont {Xiao}\ \emph {et~al.}(2021)\citenamefont {Xiao}, \citenamefont {Chen}, \citenamefont {Dong},\ and\ \citenamefont {Meng}}]{Xiao:2020ltm}%
  \BibitemOpen
  \bibfield  {author} {\bibinfo {author} {\bibfnamefont {C.-J.}\ \bibnamefont {Xiao}}, \bibinfo {author} {\bibfnamefont {D.-Y.}\ \bibnamefont {Chen}}, \bibinfo {author} {\bibfnamefont {Y.-B.}\ \bibnamefont {Dong}}, \ and\ \bibinfo {author} {\bibfnamefont {G.-W.}\ \bibnamefont {Meng}},\ }\href {\doibase 10.1103/PhysRevD.103.034004} {\bibfield  {journal} {\bibinfo  {journal} {Phys. Rev. D}\ }\textbf {\bibinfo {volume} {103}},\ \bibinfo {pages} {034004} (\bibinfo {year} {2021})},\ \Eprint {http://arxiv.org/abs/2009.14538} {arXiv:2009.14538 [hep-ph]} \BibitemShut {NoStop}%
\bibitem [{\citenamefont {Kong}\ \emph {et~al.}(2021)\citenamefont {Kong}, \citenamefont {Zhu}, \citenamefont {Song},\ and\ \citenamefont {He}}]{Kong:2021ohg}%
  \BibitemOpen
  \bibfield  {author} {\bibinfo {author} {\bibfnamefont {S.-Y.}\ \bibnamefont {Kong}}, \bibinfo {author} {\bibfnamefont {J.-T.}\ \bibnamefont {Zhu}}, \bibinfo {author} {\bibfnamefont {D.}~\bibnamefont {Song}}, \ and\ \bibinfo {author} {\bibfnamefont {J.}~\bibnamefont {He}},\ }\href {\doibase 10.1103/PhysRevD.104.094012} {\bibfield  {journal} {\bibinfo  {journal} {Phys. Rev. D}\ }\textbf {\bibinfo {volume} {104}},\ \bibinfo {pages} {094012} (\bibinfo {year} {2021})},\ \Eprint {http://arxiv.org/abs/2106.07272} {arXiv:2106.07272 [hep-ph]} \BibitemShut {NoStop}%
\bibitem [{\citenamefont {He}\ and\ \citenamefont {Chen}(2021)}]{He:2020btl}%
  \BibitemOpen
  \bibfield  {author} {\bibinfo {author} {\bibfnamefont {J.}~\bibnamefont {He}}\ and\ \bibinfo {author} {\bibfnamefont {D.-Y.}\ \bibnamefont {Chen}},\ }\href {\doibase 10.1088/1674-1137/abeda8} {\bibfield  {journal} {\bibinfo  {journal} {Chin. Phys. C}\ }\textbf {\bibinfo {volume} {45}},\ \bibinfo {pages} {063102} (\bibinfo {year} {2021})},\ \Eprint {http://arxiv.org/abs/2008.07782} {arXiv:2008.07782 [hep-ph]} \BibitemShut {NoStop}%
\bibitem [{\citenamefont {Agaev}\ \emph {et~al.}(2021)\citenamefont {Agaev}, \citenamefont {Azizi},\ and\ \citenamefont {Sundu}}]{Agaev:2020nrc}%
  \BibitemOpen
  \bibfield  {author} {\bibinfo {author} {\bibfnamefont {S.~S.}\ \bibnamefont {Agaev}}, \bibinfo {author} {\bibfnamefont {K.}~\bibnamefont {Azizi}}, \ and\ \bibinfo {author} {\bibfnamefont {H.}~\bibnamefont {Sundu}},\ }\href {\doibase 10.1088/1361-6471/ac0b31} {\bibfield  {journal} {\bibinfo  {journal} {J. Phys. G}\ }\textbf {\bibinfo {volume} {48}},\ \bibinfo {pages} {085012} (\bibinfo {year} {2021})},\ \Eprint {http://arxiv.org/abs/2008.13027} {arXiv:2008.13027 [hep-ph]} \BibitemShut {NoStop}%
\bibitem [{\citenamefont {Abreu}(2021)}]{Abreu:2020ony}%
  \BibitemOpen
  \bibfield  {author} {\bibinfo {author} {\bibfnamefont {L.~M.}\ \bibnamefont {Abreu}},\ }\href {\doibase 10.1103/PhysRevD.103.036013} {\bibfield  {journal} {\bibinfo  {journal} {Phys. Rev. D}\ }\textbf {\bibinfo {volume} {103}},\ \bibinfo {pages} {036013} (\bibinfo {year} {2021})},\ \Eprint {http://arxiv.org/abs/2010.14955} {arXiv:2010.14955 [hep-ph]} \BibitemShut {NoStop}%
\bibitem [{\citenamefont {Wang}\ and\ \citenamefont {Zhu}(2022)}]{Wang:2021lwy}%
  \BibitemOpen
  \bibfield  {author} {\bibinfo {author} {\bibfnamefont {B.}~\bibnamefont {Wang}}\ and\ \bibinfo {author} {\bibfnamefont {S.-L.}\ \bibnamefont {Zhu}},\ }\href {\doibase 10.1140/epjc/s10052-022-10396-9} {\bibfield  {journal} {\bibinfo  {journal} {Eur. Phys. J. C}\ }\textbf {\bibinfo {volume} {82}},\ \bibinfo {pages} {419} (\bibinfo {year} {2022})},\ \Eprint {http://arxiv.org/abs/2107.09275} {arXiv:2107.09275 [hep-ph]} \BibitemShut {NoStop}%
\bibitem [{\citenamefont {Ding}\ \emph {et~al.}(2024)\citenamefont {Ding}, \citenamefont {Huang},\ and\ \citenamefont {He}}]{Ding:2024dif}%
  \BibitemOpen
  \bibfield  {author} {\bibinfo {author} {\bibfnamefont {Z.-M.}\ \bibnamefont {Ding}}, \bibinfo {author} {\bibfnamefont {Q.}~\bibnamefont {Huang}}, \ and\ \bibinfo {author} {\bibfnamefont {J.}~\bibnamefont {He}},\ }\href {\doibase 10.1140/epjc/s10052-024-13214-6} {\bibfield  {journal} {\bibinfo  {journal} {Eur. Phys. J. C}\ }\textbf {\bibinfo {volume} {84}},\ \bibinfo {pages} {822} (\bibinfo {year} {2024})},\ \Eprint {http://arxiv.org/abs/2407.13503} {arXiv:2407.13503 [hep-ph]} \BibitemShut {NoStop}%
\bibitem [{\citenamefont {Agaev}\ \emph {et~al.}(2023)\citenamefont {Agaev}, \citenamefont {Azizi},\ and\ \citenamefont {Sundu}}]{Agaev:2022eyk}%
  \BibitemOpen
  \bibfield  {author} {\bibinfo {author} {\bibfnamefont {S.~S.}\ \bibnamefont {Agaev}}, \bibinfo {author} {\bibfnamefont {K.}~\bibnamefont {Azizi}}, \ and\ \bibinfo {author} {\bibfnamefont {H.}~\bibnamefont {Sundu}},\ }\href {\doibase 10.1103/PhysRevD.107.094019} {\bibfield  {journal} {\bibinfo  {journal} {Phys. Rev. D}\ }\textbf {\bibinfo {volume} {107}},\ \bibinfo {pages} {094019} (\bibinfo {year} {2023})},\ \Eprint {http://arxiv.org/abs/2212.12001} {arXiv:2212.12001 [hep-ph]} \BibitemShut {NoStop}%
\bibitem [{\citenamefont {Yu}\ \emph {et~al.}(2024)\citenamefont {Yu}, \citenamefont {Wu},\ and\ \citenamefont {Chen}}]{Yu:2023avh}%
  \BibitemOpen
  \bibfield  {author} {\bibinfo {author} {\bibfnamefont {Z.}~\bibnamefont {Yu}}, \bibinfo {author} {\bibfnamefont {Q.}~\bibnamefont {Wu}}, \ and\ \bibinfo {author} {\bibfnamefont {D.-Y.}\ \bibnamefont {Chen}},\ }\href {\doibase 10.1140/epjc/s10052-024-13366-5} {\bibfield  {journal} {\bibinfo  {journal} {Eur. Phys. J. C}\ }\textbf {\bibinfo {volume} {84}},\ \bibinfo {pages} {985} (\bibinfo {year} {2024})},\ \Eprint {http://arxiv.org/abs/2310.12398} {arXiv:2310.12398 [hep-ph]} \BibitemShut {NoStop}%
\bibitem [{\citenamefont {Karliner}\ and\ \citenamefont {Rosner}(2020)}]{Karliner:2020vsi}%
  \BibitemOpen
  \bibfield  {author} {\bibinfo {author} {\bibfnamefont {M.}~\bibnamefont {Karliner}}\ and\ \bibinfo {author} {\bibfnamefont {J.~L.}\ \bibnamefont {Rosner}},\ }\href {\doibase 10.1103/PhysRevD.102.094016} {\bibfield  {journal} {\bibinfo  {journal} {Phys. Rev. D}\ }\textbf {\bibinfo {volume} {102}},\ \bibinfo {pages} {094016} (\bibinfo {year} {2020})},\ \Eprint {http://arxiv.org/abs/2008.05993} {arXiv:2008.05993 [hep-ph]} \BibitemShut {NoStop}%
\bibitem [{\citenamefont {Yang}\ \emph {et~al.}(2021)\citenamefont {Yang}, \citenamefont {Ping},\ and\ \citenamefont {Segovia}}]{Yang:2021izl}%
  \BibitemOpen
  \bibfield  {author} {\bibinfo {author} {\bibfnamefont {G.}~\bibnamefont {Yang}}, \bibinfo {author} {\bibfnamefont {J.}~\bibnamefont {Ping}}, \ and\ \bibinfo {author} {\bibfnamefont {J.}~\bibnamefont {Segovia}},\ }\href {\doibase 10.1103/PhysRevD.103.074011} {\bibfield  {journal} {\bibinfo  {journal} {Phys. Rev. D}\ }\textbf {\bibinfo {volume} {103}},\ \bibinfo {pages} {074011} (\bibinfo {year} {2021})},\ \Eprint {http://arxiv.org/abs/2101.04933} {arXiv:2101.04933 [hep-ph]} \BibitemShut {NoStop}%
\bibitem [{\citenamefont {\"Ozdem}\ and\ \citenamefont {Azizi}(2022)}]{Ozdem:2022ydv}%
  \BibitemOpen
  \bibfield  {author} {\bibinfo {author} {\bibfnamefont {U.}~\bibnamefont {\"Ozdem}}\ and\ \bibinfo {author} {\bibfnamefont {K.}~\bibnamefont {Azizi}},\ }\href {\doibase 10.1140/epja/s10050-022-00815-6} {\bibfield  {journal} {\bibinfo  {journal} {Eur. Phys. J. A}\ }\textbf {\bibinfo {volume} {58}},\ \bibinfo {pages} {171} (\bibinfo {year} {2022})},\ \Eprint {http://arxiv.org/abs/2202.11466} {arXiv:2202.11466 [hep-ph]} \BibitemShut {NoStop}%
\bibitem [{\citenamefont {Burns}\ and\ \citenamefont {Swanson}(2021)}]{Burns:2020epm}%
  \BibitemOpen
  \bibfield  {author} {\bibinfo {author} {\bibfnamefont {T.~J.}\ \bibnamefont {Burns}}\ and\ \bibinfo {author} {\bibfnamefont {E.~S.}\ \bibnamefont {Swanson}},\ }\href {\doibase 10.1016/j.physletb.2020.136057} {\bibfield  {journal} {\bibinfo  {journal} {Phys. Lett. B}\ }\textbf {\bibinfo {volume} {813}},\ \bibinfo {pages} {136057} (\bibinfo {year} {2021})},\ \Eprint {http://arxiv.org/abs/2008.12838} {arXiv:2008.12838 [hep-ph]} \BibitemShut {NoStop}%
\bibitem [{\citenamefont {Chand}\ and\ \citenamefont {Dalitz}(1962)}]{Chand:1962ec}%
  \BibitemOpen
  \bibfield  {author} {\bibinfo {author} {\bibfnamefont {R.}~\bibnamefont {Chand}}\ and\ \bibinfo {author} {\bibfnamefont {R.~H.}\ \bibnamefont {Dalitz}},\ }\href {\doibase 10.1016/0003-4916(62)90113-6} {\bibfield  {journal} {\bibinfo  {journal} {Annals Phys.}\ }\textbf {\bibinfo {volume} {20}},\ \bibinfo {pages} {1} (\bibinfo {year} {1962})}\BibitemShut {NoStop}%
\bibitem [{\citenamefont {Barrett}\ and\ \citenamefont {Deloff}(1999)}]{Barrett:1999cw}%
  \BibitemOpen
  \bibfield  {author} {\bibinfo {author} {\bibfnamefont {R.~C.}\ \bibnamefont {Barrett}}\ and\ \bibinfo {author} {\bibfnamefont {A.}~\bibnamefont {Deloff}},\ }\href {\doibase 10.1103/PhysRevC.60.025201} {\bibfield  {journal} {\bibinfo  {journal} {Phys. Rev. C}\ }\textbf {\bibinfo {volume} {60}},\ \bibinfo {pages} {025201} (\bibinfo {year} {1999})}\BibitemShut {NoStop}%
\bibitem [{\citenamefont {Deloff}(2000)}]{Deloff:1999gc}%
  \BibitemOpen
  \bibfield  {author} {\bibinfo {author} {\bibfnamefont {A.}~\bibnamefont {Deloff}},\ }\href {\doibase 10.1103/PhysRevC.61.024004} {\bibfield  {journal} {\bibinfo  {journal} {Phys. Rev. C}\ }\textbf {\bibinfo {volume} {61}},\ \bibinfo {pages} {024004} (\bibinfo {year} {2000})}\BibitemShut {NoStop}%
\bibitem [{\citenamefont {Kamalov}\ \emph {et~al.}(2001)\citenamefont {Kamalov}, \citenamefont {Oset},\ and\ \citenamefont {Ramos}}]{Kamalov:2000iy}%
  \BibitemOpen
  \bibfield  {author} {\bibinfo {author} {\bibfnamefont {S.~S.}\ \bibnamefont {Kamalov}}, \bibinfo {author} {\bibfnamefont {E.}~\bibnamefont {Oset}}, \ and\ \bibinfo {author} {\bibfnamefont {A.}~\bibnamefont {Ramos}},\ }\href {\doibase 10.1016/S0375-9474(00)00709-0} {\bibfield  {journal} {\bibinfo  {journal} {Nucl. Phys. A}\ }\textbf {\bibinfo {volume} {690}},\ \bibinfo {pages} {494} (\bibinfo {year} {2001})},\ \Eprint {http://arxiv.org/abs/nucl-th/0010054} {arXiv:nucl-th/0010054} \BibitemShut {NoStop}%
\bibitem [{\citenamefont {Oset}\ and\ \citenamefont {Ramos}(2010)}]{Oset:2010tof}%
  \BibitemOpen
  \bibfield  {author} {\bibinfo {author} {\bibfnamefont {E.}~\bibnamefont {Oset}}\ and\ \bibinfo {author} {\bibfnamefont {A.}~\bibnamefont {Ramos}},\ }\href {\doibase 10.1140/epja/i2010-10957-3} {\bibfield  {journal} {\bibinfo  {journal} {Eur. Phys. J. A}\ }\textbf {\bibinfo {volume} {44}},\ \bibinfo {pages} {445} (\bibinfo {year} {2010})},\ \Eprint {http://arxiv.org/abs/0905.0973} {arXiv:0905.0973 [hep-ph]} \BibitemShut {NoStop}%
\bibitem [{\citenamefont {Roca}\ and\ \citenamefont {Oset}(2010)}]{Roca:2010tf}%
  \BibitemOpen
  \bibfield  {author} {\bibinfo {author} {\bibfnamefont {L.}~\bibnamefont {Roca}}\ and\ \bibinfo {author} {\bibfnamefont {E.}~\bibnamefont {Oset}},\ }\href {\doibase 10.1103/PhysRevD.82.054013} {\bibfield  {journal} {\bibinfo  {journal} {Phys. Rev. D}\ }\textbf {\bibinfo {volume} {82}},\ \bibinfo {pages} {054013} (\bibinfo {year} {2010})},\ \Eprint {http://arxiv.org/abs/1005.0283} {arXiv:1005.0283 [hep-ph]} \BibitemShut {NoStop}%
\bibitem [{\citenamefont {Xiao}\ \emph {et~al.}(2011)\citenamefont {Xiao}, \citenamefont {Bayar},\ and\ \citenamefont {Oset}}]{Xiao:2011rc}%
  \BibitemOpen
  \bibfield  {author} {\bibinfo {author} {\bibfnamefont {C.~W.}\ \bibnamefont {Xiao}}, \bibinfo {author} {\bibfnamefont {M.}~\bibnamefont {Bayar}}, \ and\ \bibinfo {author} {\bibfnamefont {E.}~\bibnamefont {Oset}},\ }\href {\doibase 10.1103/PhysRevD.84.034037} {\bibfield  {journal} {\bibinfo  {journal} {Phys. Rev. D}\ }\textbf {\bibinfo {volume} {84}},\ \bibinfo {pages} {034037} (\bibinfo {year} {2011})},\ \Eprint {http://arxiv.org/abs/1106.0459} {arXiv:1106.0459 [hep-ph]} \BibitemShut {NoStop}%
\bibitem [{\citenamefont {Ma}\ \emph {et~al.}(2023)\citenamefont {Ma} \emph {et~al.}}]{Belle:2022ywa}%
  \BibitemOpen
  \bibfield  {author} {\bibinfo {author} {\bibfnamefont {Y.}~\bibnamefont {Ma}} \emph {et~al.} (\bibinfo {collaboration} {Belle}),\ }\href {\doibase 10.1103/PhysRevLett.130.151903} {\bibfield  {journal} {\bibinfo  {journal} {Phys. Rev. Lett.}\ }\textbf {\bibinfo {volume} {130}},\ \bibinfo {pages} {151903} (\bibinfo {year} {2023})},\ \Eprint {http://arxiv.org/abs/2211.11151} {arXiv:2211.11151 [hep-ex]} \BibitemShut {NoStop}%
\bibitem [{\citenamefont {Roca}\ and\ \citenamefont {Oset}(2013)}]{Roca:2013cca}%
  \BibitemOpen
  \bibfield  {author} {\bibinfo {author} {\bibfnamefont {L.}~\bibnamefont {Roca}}\ and\ \bibinfo {author} {\bibfnamefont {E.}~\bibnamefont {Oset}},\ }\href {\doibase 10.1103/PhysRevC.88.055206} {\bibfield  {journal} {\bibinfo  {journal} {Phys. Rev. C}\ }\textbf {\bibinfo {volume} {88}},\ \bibinfo {pages} {055206} (\bibinfo {year} {2013})},\ \Eprint {http://arxiv.org/abs/1307.5752} {arXiv:1307.5752 [nucl-th]} \BibitemShut {NoStop}%
\bibitem [{\citenamefont {Yue}\ \emph {et~al.}(2024)\citenamefont {Yue}, \citenamefont {Guo},\ and\ \citenamefont {Chen}}]{Yue:2024paz}%
  \BibitemOpen
  \bibfield  {author} {\bibinfo {author} {\bibfnamefont {Z.-L.}\ \bibnamefont {Yue}}, \bibinfo {author} {\bibfnamefont {Q.-Y.}\ \bibnamefont {Guo}}, \ and\ \bibinfo {author} {\bibfnamefont {D.-Y.}\ \bibnamefont {Chen}},\ }\href {\doibase 10.1103/PhysRevD.109.094049} {\bibfield  {journal} {\bibinfo  {journal} {Phys. Rev. D}\ }\textbf {\bibinfo {volume} {109}},\ \bibinfo {pages} {094049} (\bibinfo {year} {2024})},\ \Eprint {http://arxiv.org/abs/2402.10594} {arXiv:2402.10594 [hep-ph]} \BibitemShut {NoStop}%
\bibitem [{\citenamefont {He}\ \emph {et~al.}(2007)\citenamefont {He}, \citenamefont {Li}, \citenamefont {Liu},\ and\ \citenamefont {Zeng}}]{He:2006is}%
  \BibitemOpen
  \bibfield  {author} {\bibinfo {author} {\bibfnamefont {X.-G.}\ \bibnamefont {He}}, \bibinfo {author} {\bibfnamefont {X.-Q.}\ \bibnamefont {Li}}, \bibinfo {author} {\bibfnamefont {X.}~\bibnamefont {Liu}}, \ and\ \bibinfo {author} {\bibfnamefont {X.-Q.}\ \bibnamefont {Zeng}},\ }\href {\doibase 10.1140/epjc/s10052-007-0347-y} {\bibfield  {journal} {\bibinfo  {journal} {Eur. Phys. J. C}\ }\textbf {\bibinfo {volume} {51}},\ \bibinfo {pages} {883} (\bibinfo {year} {2007})},\ \Eprint {http://arxiv.org/abs/hep-ph/0606015} {arXiv:hep-ph/0606015} \BibitemShut {NoStop}%
\bibitem [{\citenamefont {Dong}\ \emph {et~al.}(2010)\citenamefont {Dong}, \citenamefont {Faessler}, \citenamefont {Gutsche}, \citenamefont {Kumano},\ and\ \citenamefont {Lyubovitskij}}]{Dong:2010xv}%
  \BibitemOpen
  \bibfield  {author} {\bibinfo {author} {\bibfnamefont {Y.}~\bibnamefont {Dong}}, \bibinfo {author} {\bibfnamefont {A.}~\bibnamefont {Faessler}}, \bibinfo {author} {\bibfnamefont {T.}~\bibnamefont {Gutsche}}, \bibinfo {author} {\bibfnamefont {S.}~\bibnamefont {Kumano}}, \ and\ \bibinfo {author} {\bibfnamefont {V.~E.}\ \bibnamefont {Lyubovitskij}},\ }\href {\doibase 10.1103/PhysRevD.82.034035} {\bibfield  {journal} {\bibinfo  {journal} {Phys. Rev. D}\ }\textbf {\bibinfo {volume} {82}},\ \bibinfo {pages} {034035} (\bibinfo {year} {2010})},\ \Eprint {http://arxiv.org/abs/1006.4018} {arXiv:1006.4018 [hep-ph]} \BibitemShut {NoStop}%
\bibitem [{\citenamefont {He}\ \emph {et~al.}(2010)\citenamefont {He}, \citenamefont {Ye}, \citenamefont {Sun},\ and\ \citenamefont {Liu}}]{He:2010zq}%
  \BibitemOpen
  \bibfield  {author} {\bibinfo {author} {\bibfnamefont {J.}~\bibnamefont {He}}, \bibinfo {author} {\bibfnamefont {Y.-T.}\ \bibnamefont {Ye}}, \bibinfo {author} {\bibfnamefont {Z.-F.}\ \bibnamefont {Sun}}, \ and\ \bibinfo {author} {\bibfnamefont {X.}~\bibnamefont {Liu}},\ }\href {\doibase 10.1103/PhysRevD.82.114029} {\bibfield  {journal} {\bibinfo  {journal} {Phys. Rev. D}\ }\textbf {\bibinfo {volume} {82}},\ \bibinfo {pages} {114029} (\bibinfo {year} {2010})},\ \Eprint {http://arxiv.org/abs/1008.1500} {arXiv:1008.1500 [hep-ph]} \BibitemShut {NoStop}%
\bibitem [{\citenamefont {Ortega}\ \emph {et~al.}(2013)\citenamefont {Ortega}, \citenamefont {Entem},\ and\ \citenamefont {Fernandez}}]{Ortega:2013fta}%
  \BibitemOpen
  \bibfield  {author} {\bibinfo {author} {\bibfnamefont {P.~G.}\ \bibnamefont {Ortega}}, \bibinfo {author} {\bibfnamefont {D.~R.}\ \bibnamefont {Entem}}, \ and\ \bibinfo {author} {\bibfnamefont {F.}~\bibnamefont {Fernandez}},\ }\href {\doibase 10.1007/s00601-012-0569-x} {\bibfield  {journal} {\bibinfo  {journal} {Few Body Syst.}\ }\textbf {\bibinfo {volume} {54}},\ \bibinfo {pages} {1101} (\bibinfo {year} {2013})}\BibitemShut {NoStop}%
\bibitem [{\citenamefont {Wang}\ \emph {et~al.}(2020)\citenamefont {Wang}, \citenamefont {Meng},\ and\ \citenamefont {Zhu}}]{Wang:2020dhf}%
  \BibitemOpen
  \bibfield  {author} {\bibinfo {author} {\bibfnamefont {B.}~\bibnamefont {Wang}}, \bibinfo {author} {\bibfnamefont {L.}~\bibnamefont {Meng}}, \ and\ \bibinfo {author} {\bibfnamefont {S.-L.}\ \bibnamefont {Zhu}},\ }\href {\doibase 10.1103/PhysRevD.101.094035} {\bibfield  {journal} {\bibinfo  {journal} {Phys. Rev. D}\ }\textbf {\bibinfo {volume} {101}},\ \bibinfo {pages} {094035} (\bibinfo {year} {2020})},\ \Eprint {http://arxiv.org/abs/2003.05688} {arXiv:2003.05688 [hep-ph]} \BibitemShut {NoStop}%
\bibitem [{\citenamefont {Kong}\ \emph {et~al.}(2024)\citenamefont {Kong}, \citenamefont {Zhu}, \citenamefont {Chen},\ and\ \citenamefont {He}}]{Kong:2024scz}%
  \BibitemOpen
  \bibfield  {author} {\bibinfo {author} {\bibfnamefont {S.-Y.}\ \bibnamefont {Kong}}, \bibinfo {author} {\bibfnamefont {J.-T.}\ \bibnamefont {Zhu}}, \bibinfo {author} {\bibfnamefont {S.}~\bibnamefont {Chen}}, \ and\ \bibinfo {author} {\bibfnamefont {J.}~\bibnamefont {He}},\ }\href@noop {} {\  (\bibinfo {year} {2024})},\ \Eprint {http://arxiv.org/abs/2402.02703} {arXiv:2402.02703 [hep-ph]} \BibitemShut {NoStop}%
\bibitem [{\citenamefont {Weinberg}(1965)}]{PhysRev.137.B672}%
  \BibitemOpen
  \bibfield  {author} {\bibinfo {author} {\bibfnamefont {S.}~\bibnamefont {Weinberg}},\ }\href {\doibase 10.1103/PhysRev.137.B672} {\bibfield  {journal} {\bibinfo  {journal} {Phys. Rev.}\ }\textbf {\bibinfo {volume} {137}},\ \bibinfo {pages} {B672} (\bibinfo {year} {1965})}\BibitemShut {NoStop}%
\bibitem [{\citenamefont {Gamermann}\ \emph {et~al.}(2010)\citenamefont {Gamermann}, \citenamefont {Nieves}, \citenamefont {Oset},\ and\ \citenamefont {Ruiz~Arriola}}]{Gamermann:2009uq}%
  \BibitemOpen
  \bibfield  {author} {\bibinfo {author} {\bibfnamefont {D.}~\bibnamefont {Gamermann}}, \bibinfo {author} {\bibfnamefont {J.}~\bibnamefont {Nieves}}, \bibinfo {author} {\bibfnamefont {E.}~\bibnamefont {Oset}}, \ and\ \bibinfo {author} {\bibfnamefont {E.}~\bibnamefont {Ruiz~Arriola}},\ }\href {\doibase 10.1103/PhysRevD.81.014029} {\bibfield  {journal} {\bibinfo  {journal} {Phys. Rev. D}\ }\textbf {\bibinfo {volume} {81}},\ \bibinfo {pages} {014029} (\bibinfo {year} {2010})},\ \Eprint {http://arxiv.org/abs/0911.4407} {arXiv:0911.4407 [hep-ph]} \BibitemShut {NoStop}%
\bibitem [{\citenamefont {Liu}()}]{talkbeihang}%
  \BibitemOpen
  \bibfield  {author} {\bibinfo {author} {\bibfnamefont {B.}~\bibnamefont {Liu}},\ }\bibinfo {note} {, talk of Beijang Liu in the Few Body Conference (2024)}\BibitemShut {NoStop}%
\bibitem [{\citenamefont {Sekihara}\ \emph {et~al.}(2016)\citenamefont {Sekihara}, \citenamefont {Oset},\ and\ \citenamefont {Ramos}}]{Sekihara:2016vyd}%
  \BibitemOpen
  \bibfield  {author} {\bibinfo {author} {\bibfnamefont {T.}~\bibnamefont {Sekihara}}, \bibinfo {author} {\bibfnamefont {E.}~\bibnamefont {Oset}}, \ and\ \bibinfo {author} {\bibfnamefont {A.}~\bibnamefont {Ramos}},\ }\href {\doibase 10.1093/ptep/ptw166} {\bibfield  {journal} {\bibinfo  {journal} {PTEP}\ }\textbf {\bibinfo {volume} {2016}},\ \bibinfo {pages} {123D03} (\bibinfo {year} {2016})},\ \Eprint {http://arxiv.org/abs/1607.02058} {arXiv:1607.02058 [hep-ph]} \BibitemShut {NoStop}%
\bibitem [{\citenamefont {Sada}\ \emph {et~al.}(2016)\citenamefont {Sada} \emph {et~al.}}]{J-PARCE15:2016esq}%
  \BibitemOpen
  \bibfield  {author} {\bibinfo {author} {\bibfnamefont {Y.}~\bibnamefont {Sada}} \emph {et~al.} (\bibinfo {collaboration} {J-PARC E15}),\ }\href {\doibase 10.1093/ptep/ptw040} {\bibfield  {journal} {\bibinfo  {journal} {PTEP}\ }\textbf {\bibinfo {volume} {2016}},\ \bibinfo {pages} {051D01} (\bibinfo {year} {2016})},\ \Eprint {http://arxiv.org/abs/1601.06876} {arXiv:1601.06876 [nucl-ex]} \BibitemShut {NoStop}%
\end{thebibliography}%

\end{document}